**The terrestrial planet formation paradox inferred from high-resolution N-body simulations**

J. M. Y. Woo[a,b,c], R. Brasser[d], S. L. Grimm[e], M. L. Timpe[a], J. Stadel[a]

[a]Institute for Computational Science, University of Zürich, Winterthurerstrasse 190, 8057 Zürich, Switzerland
[b]Institut für Planetologie, University of Münster, Wilhelm-Klemm-Straße 10, 48149, Germany
[c]Laboratoire Lagrange, Université Cote d'Azur, CNRS,Observatoire de la Côte d'Azur, Boulevard de l'Observatoire, 06304 Nice Cedex 4, France
[d]Origins Research Institute, Research Centre for Astronomy and Earth Sciences; Budapest H-1121, Hungary
[e]Center for Space and Habitability, University of Bern, Gesellschaftsstrasse 6, 3012 Bern, Switzerland

**Abstract**

Recent improvements to GPU hardware and the symplectic N-body code *GENGA* allow for unprecedented resolution in simulations of planet formation. In this paper, we report results from high-resolution N-body simulations of terrestrial planet formation that are mostly direct continuation of our previous 10 Myr simulations (Woo et al. 2021a) until 150 Myr. By assuming that Jupiter and Saturn have always maintained their current eccentric orbits (EJS), we are able to achieve a reasonably good match to the current inner solar system architecture. However, due to the strong radial mixing that occurs in the EJS scenario, it has difficulties in explaining the known isotopic differences between bodies in the inner solar system, most notably between Earth and Mars. On the other hand, assuming initially circular orbits for Jupiter and Saturn (CJS) can reproduce the observed low degree of radial mixing in the inner solar system, while failing to reproduce the current architecture of the inner solar system. These outcomes suggest a possible paradox between dynamical structure and cosmochemical data for the terrestrial planets within the classical formation scenario. In addition, we use our high-resolution simulations to study the assembly of Earth in unprecedented detail, focusing on its collisional history. We find that more than 90% of giant impacts (GIs) on Earth occur within 80 Myr in the EJS simulations, matching the possible early timing of the Moon-forming GI based on the ages recorded by various meteoritic samples. However, the CJS and EJS scenarios both result in a leftover planetesimal mass that is 1-2 orders of magnitude greater than what currently exists in the solar system. Our overall results strongly suggest the need to consider alternative initial conditions for terrestrial planet formation in the solar system. For example, it may be necessary to assume a lower initial mass in the asteroid belt in order to lower the subsequent effect of radial mixing in the EJS scenario, as well as reduce the mass of leftover planetesimals. Similarly, in the CJS scenario, including an early giant planet instability within 80 Myr could make the scenario dynamically feasible.

1.    **Introduction**

The emergence of symplectic CPU-based N-body codes since the early 90's (e.g. Chambers, 1999; Duncan et al., 1998; Wisdom and Holman, 1991) has facilitated dynamical studies of the late stages of terrestrial planet formation. By splitting the Hamiltonian operator (the sum of the gravitational and potential energy) of the system into Keplerian drift and perturbative kick terms, these fast symplectic N-body codes can be used to model the growth of planets from their smaller building blocks. However, due to limitations on single-core performance and finite computational resources, previous studies were generally limited to simulating the growth of the terrestrial planets from an evolved protoplanetary disc inferred from runaway growth studies (e.g. Kokubo and Ida 1996, 1998, 2000, 2002). Such simulations therefore began with lunar- to Mars-sized embryos embedded in a disc of smaller planetesimals mostly more massive than Ceres (e.g. Brasser et al., 2016a; Jacobson and Morbidelli, 2014; O'Brien et al., 2006; Quintana et al., 2016; Raymond et al., 2006, 2009; Walsh et al., 2011). In these simulations planetesimals are usually treated as semi-massive particles that can only interact with and feel the gravitational force from more massive embryos, but cannot mutually interact with each other because the computational cost would be prohibitive. This approximation can significantly lower the total number of particles in the system ($N$) and keep the computational time reasonable. However, this simplified simulation setting could lead to various problems in fully understanding the process of terrestrial planet formation, especially for modelling the growth of Mars. For instance, how Mars formed quickly enough to match its isotopic chronology inferred from martian meteorites (Dauphas and Pourmand, 2011; Marchi et al., 2020; Tang and Dauphas, 2014) is difficult if Mars was mainly formed from small planetesimals instead of collisions between lunar-sized embryos (Kobayashi and Dauphas, 2013). It has also been shown that such evolved disc initial conditions result in large uncertainties in the average final bulk composition of Mars when summarizing data from hundreds of sets of N-body simulations (Woo et al., 2018). This is because Mars' final composition is always dominated by the initial composition of just a few embryos. Hence, simulations for terrestrial planet formation starting from a more primitive stage of the protoplanetary solid disc, preferably from a disc only consisting of small planetesimals (Bromley and Kenyon, 2017; Carter et al., 2015; Clement et al., 2020; Hoffmann et al., 2017; Morishima et al., 2010; Walsh and Levison, 2019), is warranted.

The development of a GPU-based N-body code provides us with an opportunity to carry out simulations of terrestrial planet formation from more primitive protoplanetary disc configurations. The GPU accelerated N-body integrator *GENGA* (Grimm and Stadel, 2014) takes advantage of the large number of computing cores in GPU cards which can perform the same instructions on multiple threads in parallel. Grimm and Stadel (2014) were able to speed up both the mutual force calculation as well as the routines for handling close encounters. The speed increase over a regular single CPU core for high $N$ is a factor of 10-30 (Grimm and Stadel, 2014). Therefore, simulations with a large number ($N > 10,000$) of fully-self gravitating objects can be solved much faster than using conventional CPU-based N-body codes.

Recent studies have adopted *GENGA* to study the formation of terrestrial planets. In a recent study (Clement et al., 2020), the authors used *GENGA* to simulate the growth of embryos in several narrow annuli from small planetesimals with $r = 100$ km ($r$ is the radii of the initial planetesimals) and then combined all annulus simulations into one single simulation after 1 Myr of evolution. Starting from 3 Myr, they simplified their simulation by lowering the resolution of the planetesimal disc in order to perform a 200 Myr simulation for terrestrial planet formation. They found that compared to previous simulations that began from only planetesimals (Carter et al., 2015; Walsh and Levison, 2019), the planets that form in the Venus-Earth region suffer fewer giant impacts. This suggests that a dynamical event, such as a giant planet instability (Tsiganis et al., 2005) is required to trigger collisions between embryos and to grow Venus and Earth to their current mass.

We also used *GENGA* (Woo et al., 2021a; here after W21a) to perform our own sets of simulations to study the growth of embryos from a disc of planetesimals with $r = 350$ km in the framework of the classical model (Chambers, 2001). In this model, Jupiter and Saturn are placed close to their current locations without migration, and the initial solid surface density follows the minimum mass Solar nebula (MMSN; Hayashi, 1981). Unlike Clement et al. (2020), we did not divide the disc into several annuli, instead we brute force calculated the entire system of ~40,000 fully interacting particles and integrated it for 10 Myr. This allows for a direct comparison of the feeding zones and the formation timescale of Mars-region embryos from our simulations with chronology and isotopic composition data. Measurements of martian meteorites suggest that Mars has finished forming within 10 Myr (Dauphas and Pourmand, 2011; Tang and Dauphas, 2014; cf. Marchi et al., 2020) and that it has a distinct composition and feeding zone from Earth (Brasser et al., 2018; Dauphas, 2017; Lodders and Fegley, 1997; Sanloup et al., 1999; Tang and Dauphas, 2014). The recent high resolution simulations (e.g. Clement et al. 2020; Walsh and Levison 2019; Woo et al. 2021a) included artificial forces to mimic the effects of the nebular gas, which is not the case in most of the "classic" low resolution simulations discussed earlier.

Our simulation results highlighted the significant effect that the initial orbits of the giant planets have on embryo growth timescales and radial mixing. Indeed, by assuming that Jupiter and Saturn maintained their current eccentric orbits (EJS), embryos in the Mars region tend to be formed faster but with a feeding zone similar to the embryos formed in the Earth region due to the implantation of asteroid belt material via sweeping secular resonances (e.g. Bromley and Kenyon, 2017; Nagasawa et al., 2000, 2005; Thommes et al., 2008). On the other hand, by assuming initially circular orbits of the giant planets (CJS), embryos in the Mars region tend to form more slowly, but their feeding zone remains distinct from those in the Earth region (Woo et al., 2021b; here after W21b) on account of weaker sweeping secular resonances. Hence, results from the EJS scenario better match the chronology of Mars, but not its suggested isotopic difference with Earth, whereas the CJS scenario results in the opposite conclusion. Combining outcomes from our previous studies, we argue for the giant planets to reside on more circular orbits during gas dissipation in W21b, but this remains to be tested in terrestrial planet formation simulations that carry on beyond 10 Myr.

From the dynamical perspective, there are still debates about the giant planet orbits during the gas disc dissipation phase. It is currently still unclear whether the giant planets possessed eccentric or circular orbits during the gas disc phase (e.g. Pierens et al., 2014). The traditional *Nice* model assumes circular orbits (CJS) for the giant planets before the following instability (Tsiganis et al., 2005), or a resonant configuration (Morbidelli et al., 2007; Nesvorný, 2011; Nesvorný and Morbidelli, 2012). However, depending on the particular structure and qualities of the gas disc, the resonant giant planets can occasionally attain moderate eccentricities (Pierens et al., 2014). Using the final orbital configuration of the gas giants from their hydrodynamical simulations, Pierens et al. (2014) claim that they successfully reproduce the outer solar system architecture in the ensuing N-body simulations. Recent N-body tests have shown that an initially eccentric Jupiter and Saturn has a higher chance of matching the excited fifth eccentric mode of today's Jupiter (Clement et al., 2021c). Also, if the giant planet's cores formed via pebble accretion (Levison et al., 2015), involving mutual scattering between planetesimals, these cores and their fully formed giant planets may not end up on very circular orbits (e < 0.01). Last, if the gas giants formed from gravitational instability of gas fragments in the protosolar disc, which could be an alternative path for giant planet formation (e.g. Boss, 1997; Vorobyov and Elbakyan, 2018), the gas giants' initial eccentricities are not well determined. Hence, we study both EJS and CJS scenarios, whereas most of the previous studies only focus on the CJS scenarios (Clement et al., 2020; Walsh and Levison, 2019).

In the past decade, different dynamical models have been developed to explain the solar system's current architecture, especially with respect to Mars' small size and the low mass of the asteroid belt (Clement et al., 2018; Izidoro et al., 2014, 2015; Raymond and Izidoro, 2017; Walsh et al., 2011), . However, none of the previous studies explicitly rule out the classical model. Indeed, some of the dynamically favourable models that include giant planet migration, for example the Grand Tack model (Walsh et al., 2011), have been shown to induce extensive mixing within the disc (Mah and Brasser, 2021; Woo et al., 2018), which could violate the isotopic constraints derived from the inner solar system meteorites (see Section 3.1 for more detail). In order to address this issue, we begin our study by performing sets of high-resolution N-body simulations with *GENGA* that are focused on the classical model.

Thanks to our high-resolution simulations, we are for the first time able to assess and compare formation scenarios with respect to their dynamical, compositional, and chronological constraints. In our previous studies (W21a,b), we focused on Mars' growth in the first 10 Myr. In this work, we first examine the final dynamical configurations resulting from the EJS and CJS scenarios from 10 Myr to the full ~100 Myr simulation (Section 3.1). We then switch our attention to Earth's growth and the timing of its last giant impact events in our simulations (Section 3.2), so as to compare to the constraints inferred from samples. Finally, taking advantage of the high fidelity of our simulations, we study the size-frequency distribution of the leftover planetesimals (Section 3.3), which has implications for the nature of the late accretion on Earth after its core formation (Chou, 1978).

## 2. Methods

We performed simulations for the classical model (Chambers, 2001) with the GPU-based N-body code *GENGA* (Grimm and Stadel, 2014). Most of these simulations are extended from the 10 Myr simulation of W21a to a total integration time of 150 Myr. Since we have already described the details of the N-body code and its implemented gas disc in W21a, we will only briefly describe the initial conditions of our simulations involved in this study. Table 1 shows the initial conditions of our simulations. The simulation sets are divided into high-resolution and low-resolution, with initial radii of the planetesimals of 350 km and 800 km, respectively. Planetesimals are distributed in between 0.5 to 3 AU. The solid surface density of the initial solid disc follows the MMSN with a profile $\Sigma_s = 7$ g cm$^{-2}$ $(a/1$ AU$)^{-3/2}$, where $a$ is the semi-major axis of the disc. The gas disc model of *GENGA* is based on Morishima et al. (2010). We assume that the gas disc decays globally with an exponential profile and with an e-folding timescale $\tau_{decay} = 1$ or 2 Myr. The initial orbits of the giant planets are set to be either having their eccentric and inclined current orbits (EJS: $a_j = 5.20$ AU, $e_j = 0.049$, $i_j = 0.33^o$ ; $a_s = 9.58$ AU, $e_s = 0.056$, $i_s = 0.93^o$, where $(a_j, e_j, i_j)$ and $(a_s, e_s, i_s)$ are the semi-major axis, eccentricity and inclination of Jupiter and Saturn, respectively), or more circular and planar orbits adopted in the Nice model (Tsiganis et al., 2005) (CJS: $a_j = 5.45$ AU, $e_j = 0.0009$, $i_j = 0^o$ ; $a_s = 8.18$ AU, $e_s = 0.0002$, $i_s = 0.5^o$).

Here we present results of our simulation from 10 to 150 Myr, which is the period of time after the gas disc dissipation and the assembly of planets through collisions between embryos. All simulations, except for the high-resolution CJS simulations, are performed until 150 Myr with a time step of 5 days. Particles are removed when they are closer than 0.1 AU from the Sun or further than 100 AU. We will demonstrate that it is not necessary to perform the high-resolution CJS simulation until 150 Myr based on the final mass-distance distribution, and the resetting ages recorded by different inner solar system meteorites which argue for the onset of giant planet migration commencing within 80 Myr of the condensation of calcium-aluminum rich inclusions (CAIs, proxy of the birth of the solar system at 4568 Ma; Bouvier and Wadhwa, 2010) (Mojzsis et al., 2019). Moreover, the particle number $N$ decreases much more slowly in CJS than in EJS, thus performing high-resolution CJS simulations is computationally challenging. For these reasons, we perform only 2 high-resolution CJS simulations to 80 Myr. We performed a total of 8 high-resolution EJS simulations to 150 Myr. Four of them are a direct continuation of the original embryo formation simulations reported in W21a. The other four are generated by redistributing the planetesimals with random nodal angles and mean anomalies at 10 Myr from the original embryo simulation.

In W21a, we also performed simulations for the depleted disc model, in which the solid surface density is depleted with respect to the MMSN beyond the Mars region (Izidoro et al., 2014; Mah and Brasser, 2021). However, we found that in this model, Mars region embryos generally form too slowly when compared to the Hf-W and Fe-Ni chronology (Dauphas and Pourmand, 2011; Tang and Dauphas, 2014). Due to the limited computational resources caused by our access to a finite number of GPUs, we decided to only focus on the classical model in the current paper. All of our simulations are performed on Nvidia Tesla V100 and P100 cards. Table 1 shows the computing time required for each of these simulations with a total computational cost exceeding 30,000 V100/P100 GPU-hours.

Table 1 - The initial conditions of our classical model simulations. We study the growth of terrestrial planets from a disc of planetesimals with different simulation set-ups, including planetesimals' radii ($r$), number of initial planetesimals ($N$), total mass of the initial disc, decay timescale of the gas disc ($\tau_{decay}$) and orbits of the giant planets - eccentric Jupiter-Saturn (EJS) and circular Jupiter-Saturn (CJS).

| | Planetes­imals' radii ($r$) | No. of initial planetes­imals ($N$) | Mass of initial disc (0.5 to 3 AU) | Gas disc dissipation timescale $\tau_{decay}$ | Giant planets' orbit | Number of simulations | Time taken per simulation (from 10 to 150 Myr) on V100/P100 card (hours) |
|---|---|---|---|---|---|---|---|
| High-resolution | 350 km | 37437 | 3.4 $M_{Earth}$ | 2 Myr | EJS | 4 | 1044.0 |
| | | | | | CJS | 2 | 3003.7[1] |
| | | | | 1 Myr | EJS | 4 | 859.0 |
| | | | | | CJS | 2 | 2882.8[1] |
| | | | | 2 Myr | EJS | 5 | 382.1 |

---

[1] High-resolution CJS simulations are only performed to 80 Myr.

| Low-resolution | 800 km | 3114 | | | CJS | 5 | 1074.2 |
| | | | | 1 Myr | EJS | 5 | 694.0 |
| | | | | | CJS | 5 | 1301.5 |
| | | | | | Total | | 30878.4 |

Note – High-resolution simulations are extended from simulation sets CL1, CL2 and CL1_T1, respectively, in Table 1 of W21a. Low resolutions simulations are extended from simulations sets CL1_LR, CL2_LR and CL1_LR_T1, respectively, in Table 1 of W21a. CJS simulations with $\tau_{decay} = 1$ Myr are newly performed for this study.

## 3. Results and Discussion

### 3.1. Mass evolution

Following W21a, we divide our system into three types of objects. We first categorise objects as *embryos, proto-embryos* and *planetesimals,* with masses $M > 1\ M_{Moon}$, $10^{-3}\ M_{Earth} < M < M_{Moon}$ and $M < 10^{-3}\ M_{Earth}$, respectively ($10^{-3}\ M_{Earth} \approx 11$ times the initial masses of planetesimals in high-resolution simulation, and more than 5 times the mass of Ceres). *Planets* with mass $> 0.05\ M_{Earth}$ (Clement et al., 2020) are included as embryos in Fig. 1 and Fig. 2. These categories of objects are defined only for data analysis purposes, and there is no difference in treatment between them during the N-body integration.

#### 3.1.1. EJS – dynamically more successful

We first stick to the conventional approach, using the final orbital architecture of the system, to evaluate whether the simulation is successful in reproducing the current solar system. Fig. 1 shows the high-resolution simulation snapshots of an EJS simulation with $\tau_{decay} = 1$ Myr starting from 10 Myr to 150 Myr. At 10 Myr, the terrestrial planets region already consists of 6 embryos with masses larger than Mars ($M_{Mars} \approx 0.11\ M_{Earth}$). Some of them exist in the Mars region, which agrees with the Hf-W and Fe-Ni chronology of forming Mars within 10 Myr (Dauphas and Pourmand, 2011; Tang and Dauphas, 2014). As we have shown in W21a, the rapid growth of embryos in this simulation can be explained in part by the implantation of asteroid belt material ($> 0.5\ M_{Earth}$) into the terrestrial planet region through the sweeping of the $\nu_6$ secular resonance of Jupiter during the gas disc dissipation (see Figure 12 of W21a). This implantation increases the solid surface density in the terrestrial planet region, which favours rapid formation of Mars-size embryos as well as the other planets. In addition, sweeping secular resonances heat up the dynamical state of the objects in the terrestrial planet region (i.e. increasing the eccentricities of the objects that the resonance swept through). Meanwhile the gas disc has almost fully dissipated at 10 Myr, after five to ten e-folding times of $\tau_{decay}$. Thus the gas drag is not sufficient to damp down the eccentricities of the proto-embryos (blue circles) and planetesimals (red circles). Hence, we observe that at 10 Myr most of the proto-embryos and planetesimals have high eccentricities ($e > 0.1$), and their orbits are crossed with the embryos (black circles), which mostly have low eccentricities ($e < 0.05$) due to the dynamical friction of the planetesimals and proto-embryos (Wetherill and Stewart, 1989). The crossing orbits between embryos, proto-embryos and planetesimals after gas disc dissipation thus favours fast growth of embryos in the terrestrial planet region.

After the gas disc dissipated, the $\nu_5$ secular resonance stays at the terrestrial planet region and keep that region dynamically hot. Due to the hot dynamical state of the system, collisions occur frequently in between 10 to 50 Myr. This depletes the total number of objects in the system. As the number of planetesimals and proto-embryos decreases through mutual collision and accretion by embryos, dynamical friction on the Mars-size embryos decreases and leads to mutual crossing of their orbits. Hence giant impacts between embryos occur more frequently after gas disc dissipation. This further decreases the total number of embryos in the terrestrial planet region. At 50 Myr, all the planets have already accreted more than 95% of their final mass. From 50 to 150 Myr, the overall mass-distance distribution of the planetary system remains the same. The planets mainly accrete objects much smaller than themselves during this period. This process continues to deplete the background leftover planetesimals and proto-embryos. According to our results most of the giant impacts (i.e. collision between two objects with mass larger than lunar mass) occur in the first 50 Myr of the simulation. We will discuss the timing of the giant impacts in more detail in the following section (Section 3.2).

From the dynamical perspective, Fig. 1 shows a relatively successful case among our simulations because it reproduces (1) Venus and Earth analogues with mass similar to the current system (~0.8 to 1 $M_{\text{Earth}}$), (2) Mars analogues with mass less than 0.3 $M_{\text{Earth}}$ and (3) a low mass excited asteroid belt with mass depleted by > 99.5 % of its original mass (see the Appendix for the average orbital statistics for all sets of simulations). Furthermore, the growth timescales of Earth and Mars in this system match the chronology. The Earth analogue finishes accreting 90% of its mass within 50 Myr and Mars finishes forming within 10 Myr, which matches their corresponding core-formation time inferred from Hf-W chronology (Dauphas and Pourmand, 2011; Kleine et al., 2009; Yin et al., 2002). The biggest problem in the orbital architecture of this simulation is Mercury's formation. We failed to form a Mercury mass planet with an excited orbit in the innermost region; instead we formed 3 planets which are all more massive than Mercury at 150 Myr. Mutual collisions and scattering between these 3 inner planets could potentially explain Mercury's low mass and high core-mass fraction, as well as its excited orbits (Chau et al., 2018, Clement et al., 2021). Mercury's formation is a long-standing problem and almost all of the previous N-body simulations have difficulties in forming Mercury analogues with its current small mass and dynamically excited orbit, even including a modified initial solid surface density within ~0.6 au (Clement and Chambers, 2021; Lykawka and Ito, 2019) or collisional fragmentation (Clement et al., 2021a, 2019; Clement and Chambers, 2021). We leave the study on Mercury's formation to a future work.

### 3.1.2.    CJS - dynamically unsuccessful

The CJS scenario, on the other hand, is less successful in reproducing the dynamical features of the current solar system. Fig. 2 shows snapshots for the high-resolution CJS simulation, but with $\tau_{\text{decay}} = 2$ Myr. The biggest difference between CJS and EJS is that the effect of the sweeping secular resonances is much weaker - or practically non-existent - in CJS simulation compared to the EJS one. Hence, the depletion of the asteroid belt through sweeping secular resonances is negligible in the CJS simulation. As a result, lunar mass embryos can form even within the outer asteroid belt within 10 Myr.

Unlike in the EJS case (Fig. 1), the growth of embryos in the CJS configuration is slower than for the EJS case because of the lower solid surface density in the terrestrial planet region and the less excited dynamical state of the system due to the much weaker sweeping secular resonance. Hence, embryos in the Mars region have difficulty reaching 1 $M_{Mars}$ after 10 Myr in CJS (W21b). From 10 to 50 Myr, all embryos remain roughly at Mars' mass in CJS, whereas in EJS, Earth-sized planets can be formed within 50 Myr. We stopped our high-resolution CJS simulation at 80 Myr and we found that it does not reproduce the current solar system's mass-distance distribution. Venus and Earth analogues have a very low mass with only about 0.2 to 0.3 $M_{Earth}$, whereas Mars analogues are sometimes even more massive than Earth and Venus analogues. There are also stranded Mars-sized embryos in the asteroid belt, which does not match with the current system either. Given the relatively cold dynamical state of the system when compared to the EJS case (see also the Appendix), the embryos are unlikely to merge together and form Earth-sized planets even if we continue the simulation to 150 Myr. In addition, the growth rate of the Earth in the CJS case is less consistent with its Hf-W core formation ages (Halliday et al., 1996; Kleine et al., 2002, 2009; Yin et al., 2002). According to the low-resolution results of the CJS that we performed until 150 Myr, the system shows a very similar orbital architecture to the high-resolution simulation in Fig. 2, and this architecture is maintained, unchanged, from 80 Myr to 150 Myr. Therefore we argue that the CJS case is inconsistent with the current architecture and growth timescale of the terrestrial planets unless the system undergoes an orbit crossing and giant impact stage by an external agent, such as an early giant planet instability (Clement et al., 2018, 2021b; Costa et al., 2020; Mojzsis et al., 2019; Nesvorný et al., 2021; Ribeiro et al., 2020).

EJS, $\tau_{decay}$ = 1 Myr, High-res.

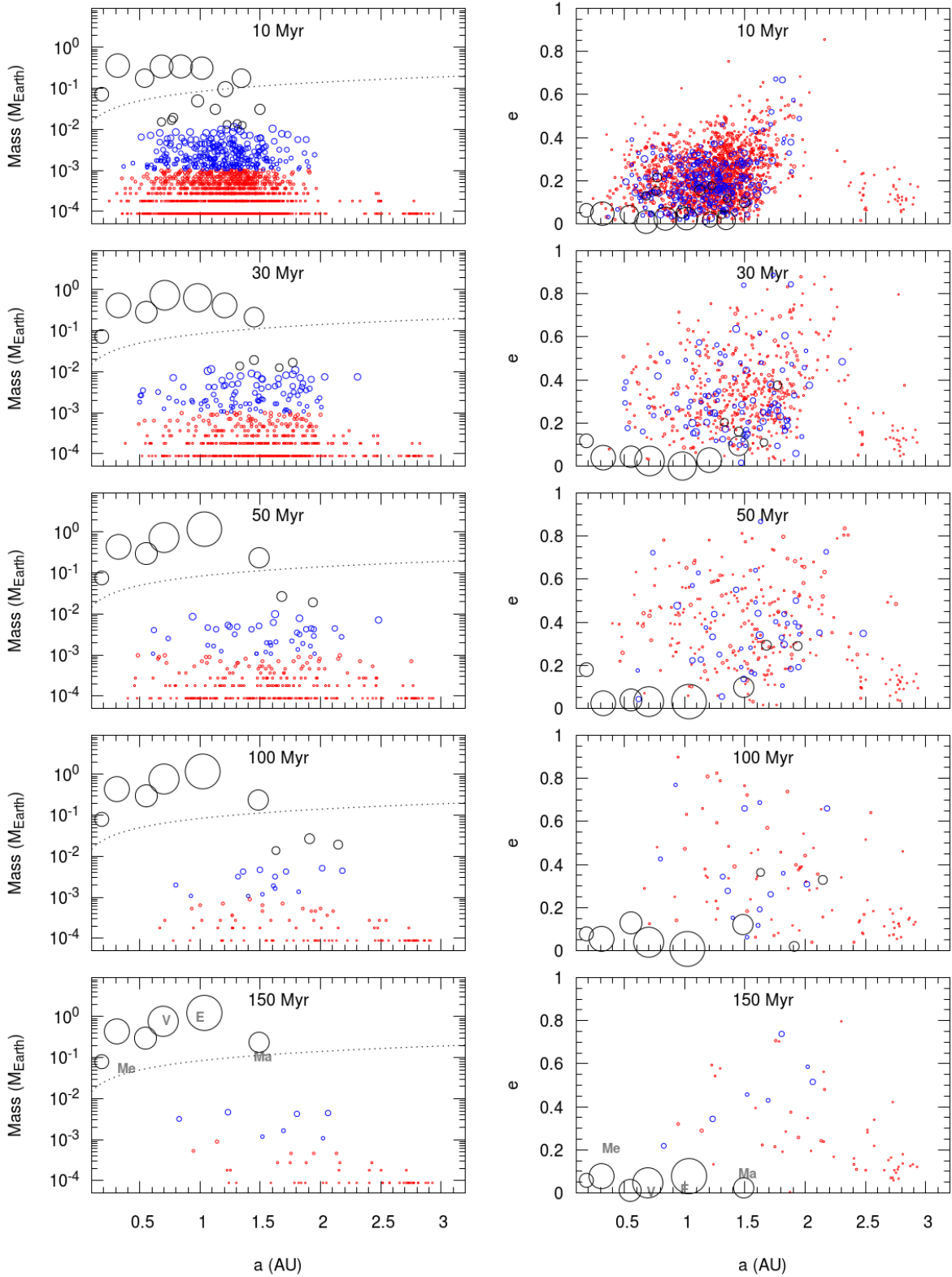

Fig. 1 - Mass versus semi-major axis (left) and eccentricity versus semi-major axis (right) of a high-resolution EJS simulation at 10, 30, 50, 100 and 150 Myr. The gas disc decay timescale is $\tau_{decay} = 1$ Myr and the initial planetesimal radius is $r = 350$ km for this simulation. The black circles represent embryos or planets with masses greater than a lunar mass ($\sim 0.012\ M_{Earth}$), blue circles represent proto-embryos with a mass less than a lunar mass but larger than $10^{-3}\ M_{Earth}$, and red circles are planetesimals with mass less than $10^{-3}\ M_{Earth}$. The bold grey letters at 150 Myr represent the current system. The dotted line on the left panels indicate the isolation mass. This system successfully reproduces the most massive planet at Venus and Earth orbits and a small Mars with less than $0.3\ M_{Earth}$, as well as an almost empty asteroid belt.

### 3.1.3. EJS vs CJS - paradox between dynamics and cosmochemistry?

To conclude, EJS is dynamically more successful than CJS in terms of reproducing the current orbital architecture of the system, as well as the Hf-W isotopic chronology of Earth and Mars. However, this does not simply imply that the giant planets must have been residing on their current orbits during the gas disc dissipation and the early assembly phase of the terrestrial planets. Instead, in W21b we argue that the giant planets more likely had more circular orbits than today, mainly because of the distinct composition of Earth and Mars inferred from isotopic measurements (e.g. $\Delta^{17}O$, $\varepsilon^{48}Ca$, $\varepsilon^{54}Cr$, $\varepsilon^{50}Ti$) on Earth's samples and martian meteorites (e.g. Dauphas et al., 2014; Franchi et al., 1999; Trinquier et al., 2009, 2007). We have shown that from the first 10 Myr of our N-body simulation, the EJS case induces strong mixing in the system that would yield Earth and Mars region embryos with highly overlapping feeding zones, whereas in the CJS case Earth and Mars region embryos have a higher chance to have almost non-overlapping feeding zones with Earth (W21b); this could have maintained a heliocentric composition gradient in the inner solar system. Such a heliocentric compositional gradient is suggested to have existed before planetesimal formation and was likely maintained during terrestrial planet assembly, as inferred from $\varepsilon^{54}Cr$ and $\varepsilon^{50}Ti$ measurements of various inner solar system meteorites (Mezger et al., 2020; Trinquier et al., 2009; Yamakawa et al., 2010). Therefore, from a cosmochemical perspective CJS is likely to be more successful than EJS.

Our results infer that either the distinct composition of Earth and Mars is a low probability outcome in the EJS simulation (which still remains to be tested in future studies), or we have to include additional dynamical effects from the gas giants to make the CJS case dynamical successful. The Nice model and **its** variants that assume CJS scenario (Tsiganis et al., 2005) could be the solution for this paradox. Recent N-body simulations imply that such an instability could occur in less than 10 Myr after the gas disc dissipates (Ribeiro et al., 2020) to avoid over growing Mars by clearing the excessive mass in the Mars and the asteroid belt region (Clement et al., 2018). This is contrary to delaying such an event for several hundreds of Myr to explain the late heavy bombardment event (Gomes et al., 2005), a scenario which has recently drawn skepticism (Mojzsis et al., 2019). The early giant planet instability also agrees with the U-Pb isotopic-resetting ages of chondritic and achondritic meteorites in different minerals with different closure temperature, which all cease before 80 Myr after the formation of the solar system (Mojzsis et al., 2019). Our CJS simulation in Fig. 2 also supports an instability within 80 Myr, or even well before 50 Myr because we form overly massive Mars analogues by 50 Myr.

Combining the high-resolution CJS simulations with the early giant planet instability is beyond the scope of our current work. In this paper, we only study the classical model in which the giant planets do not migrate and stay close to their current orbits. Given the more successful dynamical results from the EJS, we will focus our analysis on the EJS case, with the low-resolution CJS simulation as a comparison in our following discussion related to giant impact timing and the late accretion on Earth.

CJS, $\tau_{decay}$ = 2 Myr, High-res.

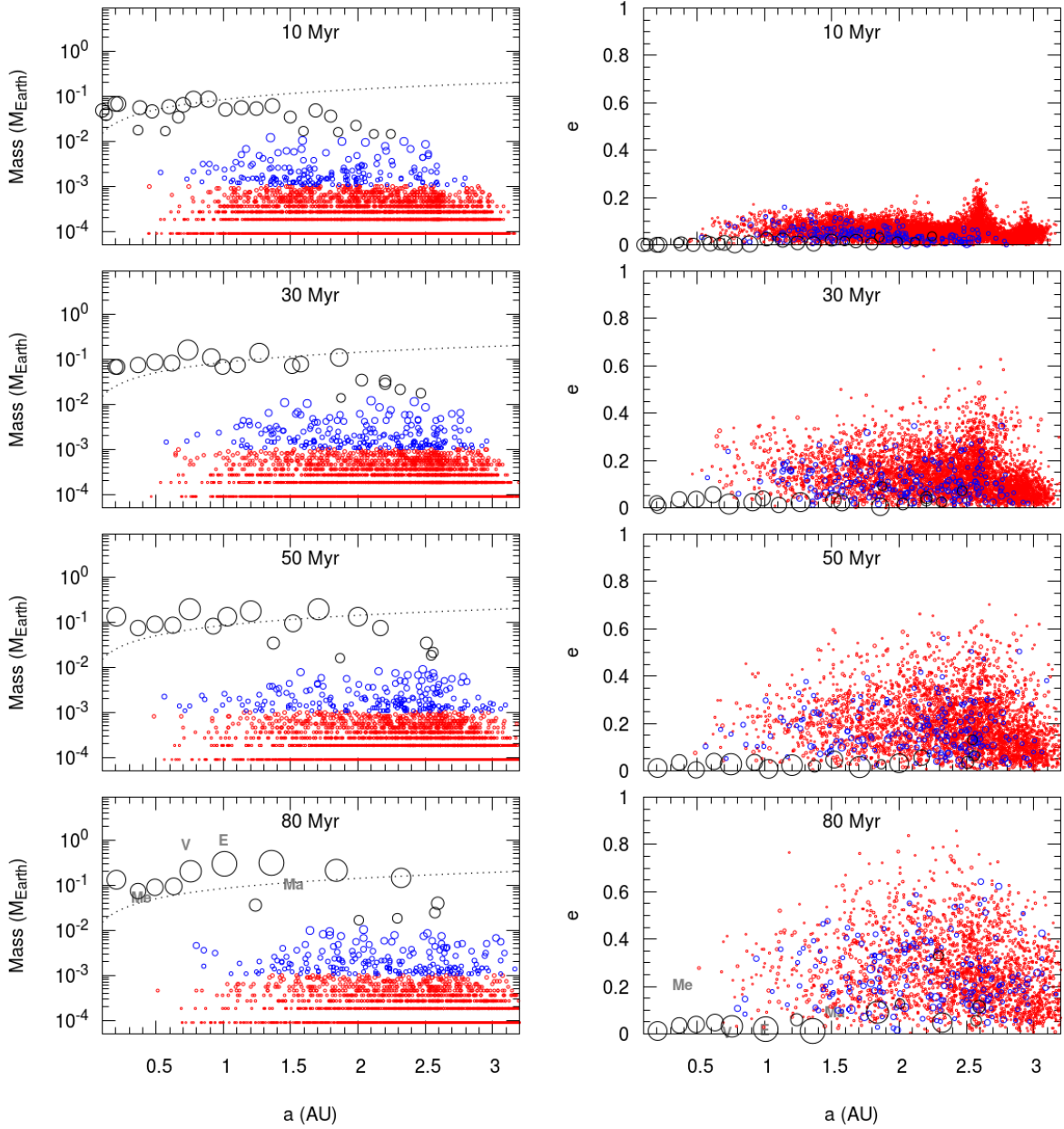

Fig. 2 - Same as Fig. 1., but for CJS with $\tau_{decay}$ = 2 Myr. We only integrate the CJS high-resolution simulations to 80 Myr. Note that this simulation fails to reproduce the architecture of the inner solar system as it yields a Mars analogue larger than Venus and Earth analogues.

## 3.2. Growth of Earth and the timing of giant impacts (GIs)

In this section, we demonstrate how simulations with different initial conditions lead to different accretion histories and GI event timings for similar planets. With respect to the collisional histories, we pay special attention to the timing of the GI on Earth analogues because there are timing constraints from both terrestrial and lunar samples. We caution the reader that the Moon-forming GI might not have been the last GI on Earth (Brasser et al., 2016b; Genda et al., 2017a).

### 3.2.1. Constraints on the timing of giant impacts on Earth

The timing of the GIs on Earth is an interesting topic because of the existence of the Moon. The Moon is thought to have formed from a GI (e.g. Benz et al., 1986; Cameron and Ward, 1976; Canup and Asphaug, 2001; Hartmann and Davis, 1975; Hosono et al., 2019; Kaib and Cowan, 2015) between the proto-Earth and a Mars-sized impactor dubbed "Theia" (Halliday, 2000). One of the big questions about the Moon-forming GI is its timing. There are still fierce debates of the timing of the event, however, we can broadly identify two scenarios, *early* and *late* Moon-forming GIs.

It has been suggested that the Moon's formation could take place relatively late, meaning from 100-200 Myr after CAIs. This is supported by the published lunar Sm-Nd ages, which generally range from 4190 Ma to 4420 Ma. In general it is thought that Sm-Nd fractionation is the result of silicate differentiation, magma ocean crystallisation and possible mantle overturn events (Borg et al., 2011, 2015, 2019, 2020; Boyet et al., 2015; Carlson et al., 2014; Gaffney and Borg, 2014; Maurice et al., 2020). The published lunar U-Pb zircon crystallisation ages also fall within the above time span (Hopkins and Mojzsis, 2015; Taylor et al., 2009), with the oldest known lunar zircon having an absolute age of 4417 Ma (Nemchin et al., 2009). These ages suggest that the Moon was still in a (global) magma ocean phase until roughly 4300 Ma, which is after the terrestrial planets completed their formation. Objects the size of the Moon are expected to cool down rapidly from the global magma ocean phase (in an order of 1000 years for solidifying 80% volume; Elkins-Tanton et al., 2011). Therefore, the proposed timing of the magma ocean crystallisation and the silicate differentiation for Moon support a late Moon-forming GI event, having occurred > 100 Myr after CAIs.

However, such a late age for magma ocean crystallisation and silicate differentiation are at odds with the inferred Lu-Hf intercept ages of the formation of first lunar crust, which range from 4480 Ma to 4510 Ma, some 50 to 80 Myr after CAIs (Barboni et al., 2017; Taylor et al., 2009). This older age of the Moon is also supported by the revised Hf-W age for lunar core formation (Thiemens et al., 2019) as well as the Pu-I-Xe age of the Moon (Swindle et al., 1986). The formation of the lunar core and its first crust cannot have occurred earlier than the formation of the Moon itself. Hence, in our interpretation of the data the measured crustal Lu-Hf formation and the Hf-W core formation age of the Moon support the earlier Moon-forming GI which occurred some 50-80 Myr after CAIs, while the Sm-Nd ages and U-Pb zircon ages most likely measure silicate differentiation, magma ocean crystallisation and additional crust formation, which tend to occur between 80 Myr and 250 Myr after CAIs.

We can also deduce the timing of the Moon-forming GI from the chronological records of Earth itself. An energetic impact of a Mars-sized object colliding with the proto-Earth would inevitably trigger whole-scale crustal melting or a core-formation and silicate differentiation event, thus resetting all the timing recorded by the isotopic decaying systems. Intriguingly, the U-Pb age of silicate differentiation (Albarede and Martine, 1984; Allègre et al., 2008; Manhes et al., 1979), I-Pu-Xe atmosphere retention age (Avice and Marty, 2014; Caracausi et al., 2016; Mukhopadhyay, 2012; Ozima and Podosek, 1999; Staudacher and Allègre, 1982) and Hf-W age of Earth's core formation (Halliday et al., 1996; Kleine et al., 2002, 2009; Yin et al., 2002) all seem to occur within 80 Myr after the formation of CAIs. This can be understood to mean that the last large impact on Earth likely occurred within ~80 Myr of CAIs, hence supporting an early Moon-forming impact. This age also coincides with the lack of crustal resetting ages after ~100 Myr of CAIs for the most resilient chronometers (e.g. U-Pb, Pb-Pb) from Mars, 4 Vesta, and various meteoritic samples from the inner solar system (Mojzsis et al., 2019). This implies that the inner solar system likely entered a quiescent stage after ~100 Myr of CAIs. Hence, combining the evidence we currently have, the last impact on Earth large enough to reset its chronometers and account for the Moon-forming GI likely occurred within 80 Myr of CAIs. The long-lived lunar magma ocean could then be attributed to the strong tidal heating experienced by the Moon in its early age (Elkins-Tanton et al., 2011).

While it is currently impossible to completely settle the debate related to the timing of the last impact and the Moon-forming GI on Earth, we take 80 Myr after CAIs as the final occurring time of any GIs on Earth in this paper's discussion based on universal terrestrial, lunar and meteoritic evidence explained above. In the following sections, we compare our simulation results with this proposed timing of the final large terrestrial impact in order to demonstrate that our simulations support an early decline in GIs on Earth.

### 3.2.2. Growth of Earth in EJS

We first demonstrate how the Earth analogues grow in our simulations. Earth analogues are defined as any planets (with mass > 0.05 $M_{Earth}$; Clement et al., 2020) with 0.85 < $a$ < 1.25 AU at 150 Myr, where $a$ is the semi-major axis of the planet. As mentioned in Section 3.1., we focus on EJS simulations because they are dynamically more successful, especially in forming Earth-size planets at 1 AU (see also the Appendix). Fig. 3 depicts the mass evolution (normalised to $M_{Earth}$) of the Earth analogues from all EJS simulations from 10 to 150 Myr. It is obvious that the probability of forming an Earth-mass planet at ~1 AU is higher in the high-resolution simulation than in the low-resolution simulation. We form four Earth analogues with ~1 $M_{Earth}$ or above in eight high-resolution simulations, while none of the Earth analogues in ten low-resolution simulations has mass > 1 $M_{Earth}$. As we have shown in W21a, the growth timescale of embryos is shorter in the low-resolution simulations because fewer collisions are required to grow to a certain mass when the initial planetesimals are larger. When the embryos grow faster, more of them are lost to the Sun due to inward Type-I migration (Tanaka et al., 2002; Tanaka and Ward, 2004). Due to the higher mass loss caused by Type-I migration, it is more difficult to form an Earth mass planet at 1 AU in the low-resolution simulations than in the high-resolution simulations. We do not show the growth of Earth analogues in the CJS simulations because we do not form any planet close to an Earth mass at 1 AU in the CJS simulations by 80 Myr (see Fig. 2 and the Appendix).

Comparing the high-resolution simulations with different gas disc decay timescales, $\tau_{\text{decay}}$, Earth analogues in simulations with a shorter $\tau_{\text{decay}}$ begin as more massive embryos at 10 Myr. Two of them have already reached ~0.7 $M_{\text{Earth}}$ at 10 Myr when $\tau_{\text{decay}} = 1$ Myr, whereas all the Earth analogues begin as Mars-sized embryos when $\tau_{\text{decay}} = 2$ Myr. This is due to the earlier implantation of asteroid belt material into the terrestrial planet region if the gas disc's lifetime is shorter. In contrast to the effect of lower resolution, in this case the early rapid embryo formation results in larger Earth analogues, since the effects of Type-I migration are also reduced.

An intriguing feature is that even though we perform our simulation for 150 Myr, most of the Earth analogues do not take that long to grow. Most of them have accreted 90% of their mass by 80 Myr (hypothesised last GI time on Earth; vertical dashed line in Fig. 3). The growth rate of these Earth analogues is fastest within 50 Myr. Most of the Earth analogues suffered a GI during their growth, which are indicated by a big jump in their mass in Fig. 3. Most (> 50%) of these jumps occur within 50 Myr. For example, in the dynamically successful case that we have shown in Fig. 1, the Earth analogue grows from ~0.7 $M_{\text{Earth}}$ to ~1.15 $M_{\text{Earth}}$ right before 50 Myr (thick red line in the top right panel of Fig. 3). The scale of this impact, with an impactor-to-target mass ratio of $\gamma \sim 0.65$, is large enough to be the Moon-forming giant impact. While a mass ratio of $\gamma \sim 0.1$ is suggested in the canonical Moon-forming scenario (Canup & Asphaug 2001), more recent studies suggest that other impact scenarios with larger values of $\gamma$, up to and including $\gamma = 1$ (Canup 2012), are also viable lunar formation scenarios.

To summarize, according to our high-resolution simulations, the primary accretion of Earth analogues finished mostly within 80 Myr. Most of the GIs suffered by the Earth analogues happen within 50 Myr, in particular those large enough to be the Moon-forming GI. This matches the recorded GI likely having occurred within 80 Myr of CAIs as discussed in the previous section.

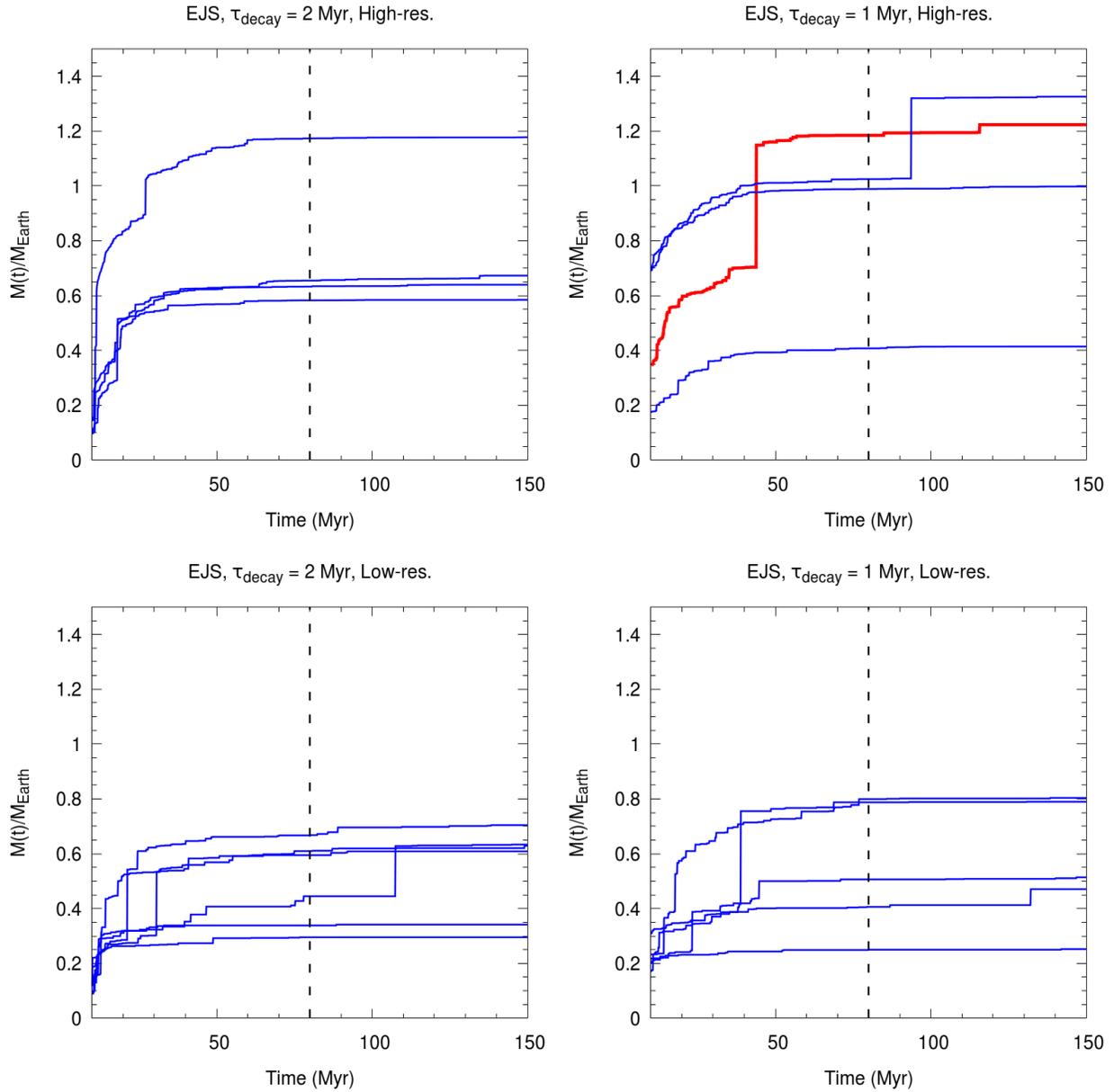

Fig. 3 - Mass growth of Earth analogues from 10 to 150 Myr in different EJS simulations. The upper and lower panels are results for high-resolution simulations (initial planetesimals' $r = 350$ km) and low-resolution simulation (initial planetesimals' $r = 800$ km), respectively. The left and right panels are results for gas dissipation timescale $\tau_{decay} = 2$ Myr and 1 Myr, respectively. The vertical black dashed line indicates the final GI time on Earth as inferred from universal samples' record (see Section 3.2.1 for the detailed discussion). The thick red line represents the growth of the Earth analogue in our dynamical successful case in Fig. 1. Most of the Earth analogue have accreted more than 90% of their final mass at 50 Myr and the higher resolution simulation has a higher chance of yielding massive Earth analogues.

### 3.2.3. Timing of giant impacts on Earth analogues in our simulations

To further quantify the timing of GIs, we perform a detailed statistical analysis, in which we demonstrate that the GIs on Earth mostly occurred within 80 Myr. Fig. 4 shows the cumulative distribution function (CDF) of the timing of giant impacts on the Earth analogues in each simulation set. A GI is defined as a collision between two embryos, both with at least a lunar mass. It is clearly demonstrated that more than 90% of the GIs occur before 80 Myr in all EJS simulations. This 90% limit can be even pushed back further to 50 Myr if we only consider data from the high-resolution simulations. Note that we only consider GIs after 10 Myr because some of our initial conditions are generated based on earlier simulation results for the first 10 Myr (see the method section). If we include those GIs within 10 Myr, then a higher fraction of GIs occur within 50 Myr. Based on our simulation results this shows that Earth has a much higher probability to suffer a GI during the earliest time of the solar system history (Yu and Jacobsen, 2011).

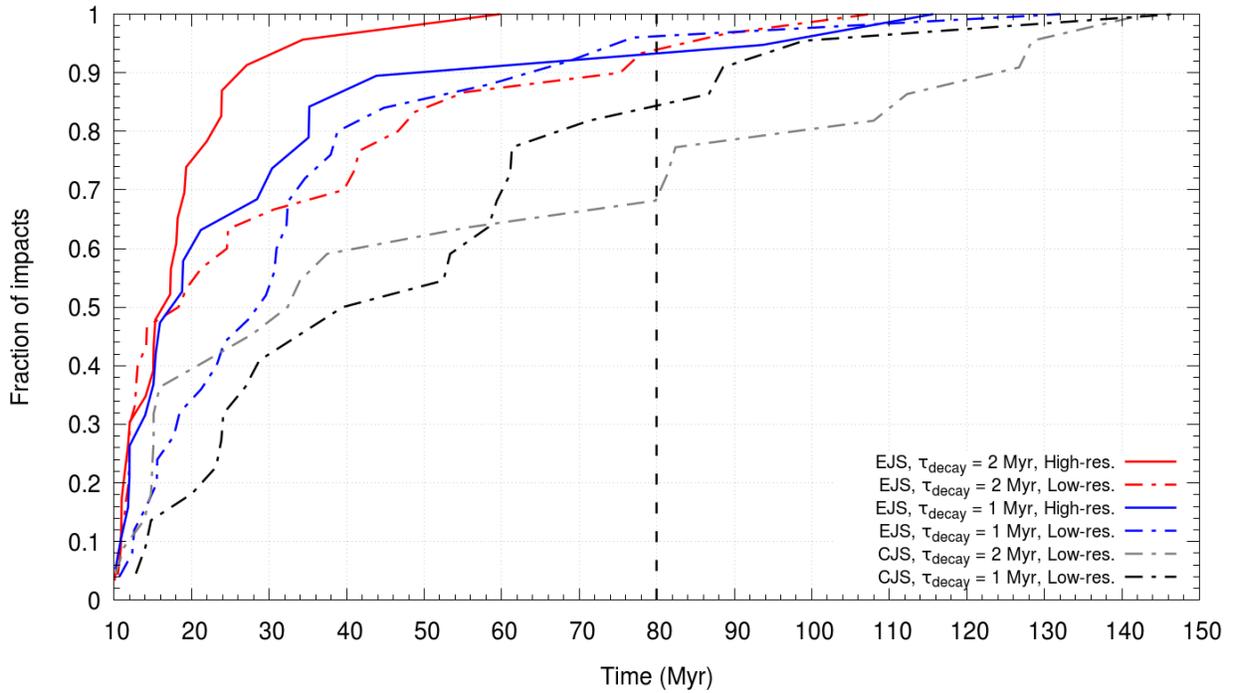

Fig. 4 - Cumulative distribution function showing the distribution of giant impact (GI) timings occuring on Earth analogues. A GI is defined as a collision between two objects with masses larger than lunar mass. The red lines are data from EJS simulation with gas dissipation timescale $\tau_{decay} = 2$ Myr, blue lines are data from EJS, $\tau_{decay} = 1$ Myr simulation; black and grey lines are data from CJS simulation with $\tau_{decay} = 2$ Myr and 1 Myr, respectively. The solid and dashed-dotted lines are data for high-resolution simulation ($r = 350$ km) and low-resolution ($r = 800$ km), respectively. The vertical black dashed line indicates the final GI time on Earth as inferred from the terrestrial, lunar and other inner solar system samples' record (see Section 3.2.1 for the detailed discussion). It is clear that most of the GIs occur before 80 Myr.

We also made a comparison between the EJS and the CJS low-resolution simulations. The CJS simulations show that a lower percentage of GIs on Earth analogues occur early on. Only ~50% to 60% of giant impacts occur within 50 Myr, compared to > 80% in the EJS simulations. This can be explained by the slower rate of accretion in the CJS simulations. As shown in Fig. 1 and Fig. 2, due to the slower growth of embryos in the CJS scenario, the total number of embryos at a particular time (e.g. at 50 Myr), is higher in the CJS simulations (Fig. 2) than in the EJS simulations (Fig. 1). The higher number of embryos at a later stage of the CJS simulation allows GIs to occur at a later time. Therefore, the EJS scenario is in general more successful in explaining the early timing of the Moon-forming GI than the CJS scenario. Nevertheless, this picture could be changed if we include an early giant planet instability before 80 Myr (Clement et al., 2018, 2021b; Costa et al., 2020; Mojzsis et al., 2019; Nesvorný et al., 2021; Ribeiro et al., 2020) into our CJS simulations, since the giant planet instability could potentially excite the terrestrial planet region, allowing mutual collisions between embryos to occur earlier. We leave such an investigation to a future study.

The hypothesis that the Moon-forming GI should occur early, within 80 Myr of CAIs is also supported by the timing of the final GI on the Earth analogues recorded in our simulations. Fig. 5 shows the mass of the impactor ((a) to (c)) and the impactor to target ratio ($\gamma$; (d) to (f)) versus GIs' occurring time of all the Earth analogues formed in each simulation sets. Due to the stochastic effect of the N-body simulations, the last giant impacts (filled-squares and filled-triangles) for the Earth analogues can span from ~10 to > 100 Myr in the EJS simulations. However, more than 70% of the Earth analogues (13 of 18) suffer a final GI within 80 Myr. In particular, all the Earth analogues in the high-resolution EJS simulation with $\tau_{decay} = 2$ Myr (filled-squares in panel (a)) suffer a final GI within ~60 Myr. We do not have any simple explanation for their peculiarly early occurrence of the final GI. It could be caused simply by the limited data set, since we have only performed 4 high-resolution simulations for each EJS set due to limits on computational resources (see Table 1).

Lowering the resolution of the simulation tends to delay the final GI time, as well as the average occurring time of all GIs, because more lunar-sized embryos can be formed when the initial planetesimal size is larger. Hence more embryos remain in the system at a later time, allowing GIs to occur at a later time. The same reason can also explain why GIs tend to occur later in the CJS simulations. Notably, the dynamical friction is weaker in the low-resolution simulations due to the lower number of planetesimals. This leads to a more dynamically excited system (see also the Appendix) and thus a higher chance of suffering embryo-embryo collisions late in the low-resolution simulations.

It has been suggested that the Moon-forming GI may not be the final GI suffered by Earth. The excess of highly siderophile elements (HSEs; meaning "iron loving") in lunar samples suggests that the Moon accreted about ~0.02 to 0.05 wt% of chondritic material after its core formation ceased (Day and Walker, 2015; Kruijer et al., 2015). In contrast, Earth accreted a much higher portion of ~0.3 to 0.8 wt% (Day et al., 2016; Walker, 2009). The accretion of chondritic material after core formation of a planet is dubbed "*late accretion*" (or "*the late veneer*"). This suggests that Earth has accreted about $1950 \pm 650$ times more mass than the Moon during late accretion, which is two orders of magnitude higher than their difference in gravitational cross-section (e.g. Bottke et al., 2010; Walker, 2009). It has been suggested that a few leftover planetesimals with the size of Ceres delivered most of the HSEs to Earth during the late accretion stage (Bottke et al., 2010). These large objects would preferentially be accreted by Earth (Sleep et al., 1989) and thus naturally explain the high contrast in HSE abundance in the terrestrial and lunar mantle. Later studies have taken this hypothesis further to suggest that the HSEs on Earth are mainly delivered by a single impactor with lunar mass (Brasser et al., 2016b; Genda et al., 2017a). This *late* *accretion GI*, occurring after the Moon-forming GI but within 80 Myr, could potentially be the final GI suffered by the Earth and mark the end of Earth's main accretion.

If the late accretion GI was in fact the last GI on Earth, then the Moon-forming GI could have been the second-to-last GI on Earth and would have occurred earlier than previously thought. The hollow squares and triangles in Fig. 5 shows the timing of the second-to-last GI that occurred on the Earth analogues in different simulation sets. Nearly 90% of the second-to-last GIs (16 of 18) in EJS simulations occur within 50 Myr, with the latest occurring no later than 80 Myr. This shows that if the Moon-forming GI is not the last GI on Earth, it probably occurs much earlier than the 4480 Ma limit inferred by the universal ages as we discussed in Section 3.3.1. Some recent studies have suggested that the formation of the lunar core and crust occurred as early as ~50 Myr and ~60 Myr, respectively, after the birth of the solar system (Barboni et al., 2017; Thiemens et al., 2019). This argues for an extremely early Moon-forming GI (Avice and Marty, 2014), which matches the second-to-last GI time recorded in our simulations.

The masses of the impactors of the final GIs from our simulations are typically smaller than the advocated mass of Theia, which is at least a Mars size object (mass > 0.1 $M_{Earth}$) with impactor-to-target mass ratio $\gamma > 0.1$ (e.g. Canup & Asphaug 2001; Canup 2012). Only 3 out of 18 impactors of the final GIs has masses > 0.1 $M_{Earth}$ in the EJS simulations (Fig. 5). Most of the impactors (12 out of 18) have masses < 0.05 $M_{Earth}$ and $\gamma < 0.1$. Compared to the final GIs, the second-to-last GIs in general is larger. Even though the impactors of the second-to-last GIs are still mostly less massive than Mars, more than half of them have masses > 0.05 $M_{Earth}$ and $\gamma > 0.1$. Compared to EJS simulations, more than half of the second-to-last and final GIs in CJS simulations has $\gamma > 0.1$, but with impactor's masses < 0.1 $M_{Earth}$. This is because almost all Earth analogues in our CJS simulations failed to grow over 0.5 $M_{Earth}$ and the last two GIs of these "fail" Earth analogues are mostly collisions between sub-martian mass embryos with martian-mass embryos.

To summarize, we find that the final giant impacts on Earth analogues in our simulations tend to occur within 80 Myr and therefore support an early Moon-forming impact scenario. Furthermore, if the Moon-forming GI is *not* responsible for the delivery of HSEs to the terrestrial mantle (i.e., it is not the final GI suffered by the Earth), then the Moon-forming GI is likely the second-to-last GI suffered by Earth and could occur as early as 20-30 Myr after CAI formation. The masses of the impactors of the second-to-last GIs, as well as their $\gamma$, from our simulations are also larger and thus closer to the suggested mass of Theia and the scale of the Moon-forming GI.

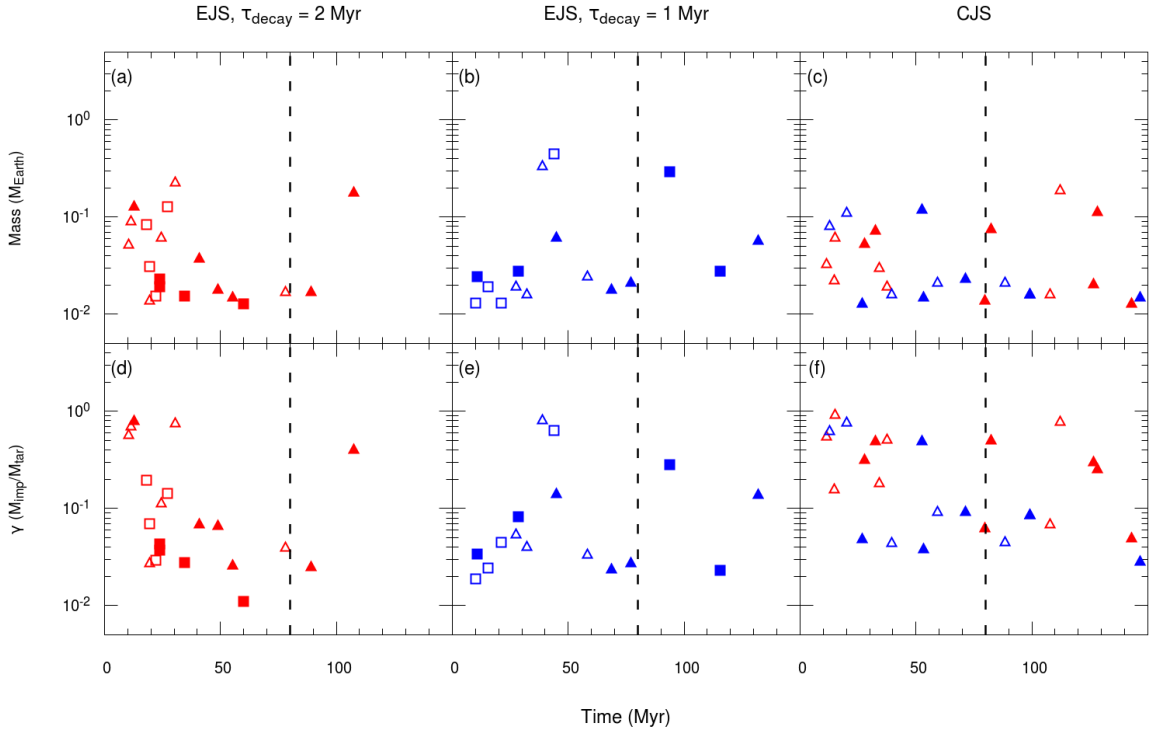

Fig. 5 – Masses of the impactors ((a) to (c)) and the impactor to target ratio (γ; (d) to (f)) versus GI occurrences times of all the Earth analogues formed in each simulation sets. Hollow data points represent the second-to-last giant impacts suffered by each Earth analogues, whereas filled data points represents the final giant impacts suffered by each Earth analogues. Squares are data from high-resolution simulations ($r$ = 350 km) and triangles are data from low-resolution simulations ($r$ = 800 km). Red points are data from simulation with a gas dissipation timescale of $\tau_{decay}$ = 2 Myr, while blue points are data from simulations with $\tau_{decay}$ = 1 Myr. The vertical black dashed line indicates the final GI time on Earth as inferred from universal samples' record (see Section 3.2.1 for the detailed discussion). Note that 100% of the second-to-last GIs and > 70% (13 of 18) of the final GIs in the EJS simulations occur ≦ 80 Myr. In general, the impact scale and impactors' masses of the second-to-last GIs is larger than the final GIs.

### 3.3. Condition for leftover planetesimals for late accretion

We have shown that the penultimate GI on Earth, which could be the Moon-forming GI, could have happened before 50 Myr. If the Moon-forming GI is the last energetic event to trigger the iron-silicate separation of Earth (i.e., core formation), its occurrence denotes the beginning of the late accretion stage on Earth. After the Moon-forming GI, any impactors that collide with Earth and successfully equilibrate with the terrestrial mantle can contribute to the excess HSE abundances measured from the terrestrial samples (Rubie et al., 2011, 2015). A lunar-size impactor (dubbed "Moneta"; Genda et al., 2017a) with ~1 wt% of Earth should be sufficient to deliver most of the excess HSEs to the terrestrial mantle (Brasser et al., 2016b, 2020; Genda et al., 2017a), although it has been proposed that the Moon-forming event was the source of the same HSEs (Sleep, 2016) in which case this event could have occurred around 4480 Ma. An interesting question that arises would be how many such Moneta-size impactors are still present in our simulations at the beginning of the late accretion phase.

According to Fig. 5, the timing of the second-to-last GI is mostly before 50 Myr in all sets of simulations. We therefore examine the size-frequency distribution (SFD) of the leftover planetesimals at 50 Myr. Fig. 6 depicts the SFD of the objects with radii ($R$) less than 2000 km in the EJS simulations at 50 Myr. Our high-resolution simulations on average form about 10-15 Moneta-sized objects (i.e. lunar-sized) per simulation, with $R$ in between 1500 km and 2000 km. Similar amounts of Moneta-sized objects are formed in the low-resolution simulations. Hence, our simulations naturally form Moneta-sized objects during the late accretion phase for Earth. Similarly, we also form ~100 Ceres-sized objects, one of which could be responsible for late accretion to Mars, the excavation of the Borealis basin and the formation of the martian satellites (Bottke and Andrews-Hanna, 2017; Bouvier et al., 2018; Brasser et al., 2017; Canup and Salmon, 2018; Citron et al., 2015; Craddock, 2011; Marinova et al., 2008; Rosenblatt et al., 2016; Woo et al., 2019).

However, one feature that we recognise from the high-resolution SFD in Fig. 6 is that the slope of the SFD becomes much steeper for $R > 1000$ km. This indicates a stark drop in the total number of Moneta-sized to sub-Moneta-sized objects and thus most of the mass of the leftover planetesimals is expected to be concentrated in objects with diameters much smaller than Moneta ($R < 1000$ km). Table 2 shows the percentages of mass distribution for objects with radii less than 2000 km at 50 Myr. Due to the drop in the number of Moneta-sized objects, they only contribute ~25 % of the leftover planetesimal mass. On the other hand, sub-Moneta-sized to Ceres-sized objects ($R$ between 500 and 1500 km) compose more than 60% of the total leftover planetesimal mass at 50 Myr. Results for the low-resolution simulations are similar to the high-resolution simulations in the bins where data is available.

Our leftover planetesimals' SFD is different from the one proposed by Bottke et al. (2010), in which they suggested that the leftover planetesimals' SFD remains shallow even for objects close to the diameter of Moneta. Bottke et al. (2010) argue that such an SFD would naturally explain the extremely high Earth-to-Moon HSE abundance ratio, because Earth's late accretion would be dominated by several large impactors which prefer to collide with Earth but not with the Moon (see also Sleep et al., 1989). However, this is not the case in our simulations. The steeper SFD at Moneta-size objects from our simulations do not favour late accretion for Earth to be delivered by only one single Moneta-sized object, as Earth is very likely to have accreted a significant amount of mass in leftover planetesimals with diameters smaller than Moneta after the second-to-last GI because more mass resides in these smaller impactors. These smaller impactors could also contribute to the delivery of the excess HSEs to the terrestrial, martian and lunar mantles.

The leftover planetesimal' mass at 50 Myr is about ~0.20 to 0.25 $M_{\text{Earth}}$ (see Table 2). This number is 1-2 orders of magnitude higher than previous estimations using the constraint imposed by the high HSE abundances of Earth and Moon (Brasser et al., 2016b, 2020; Raymond et al., 2013). Hence, our simulation results would likely overshoot both the terrestrial and lunar HSE abundances, given that we also have a significant amount of this leftover mass (~8 %, equivalent to ~0.02 $M_{\text{Earth}}$) in objects with R < 500 km that would have been accreted excessively. One solution is to assume a lower initial mass in the asteroid belt (Izidoro et al., 2014, 2015; Mah and Brasser, 2021; Raymond and Izidoro, 2017), so that the sweeping secular resonances would not be able to implant too much asteroid belt material (> 0.5 $M_{\text{Earth}}$) to the terrestrial planet region. Although this may impose difficulties in forming Mars quickly enough (W21b), lowering the mass of the original asteroid belt could slightly mitigate the radial mixing effect caused by the sweeping secular resonance (W21a).

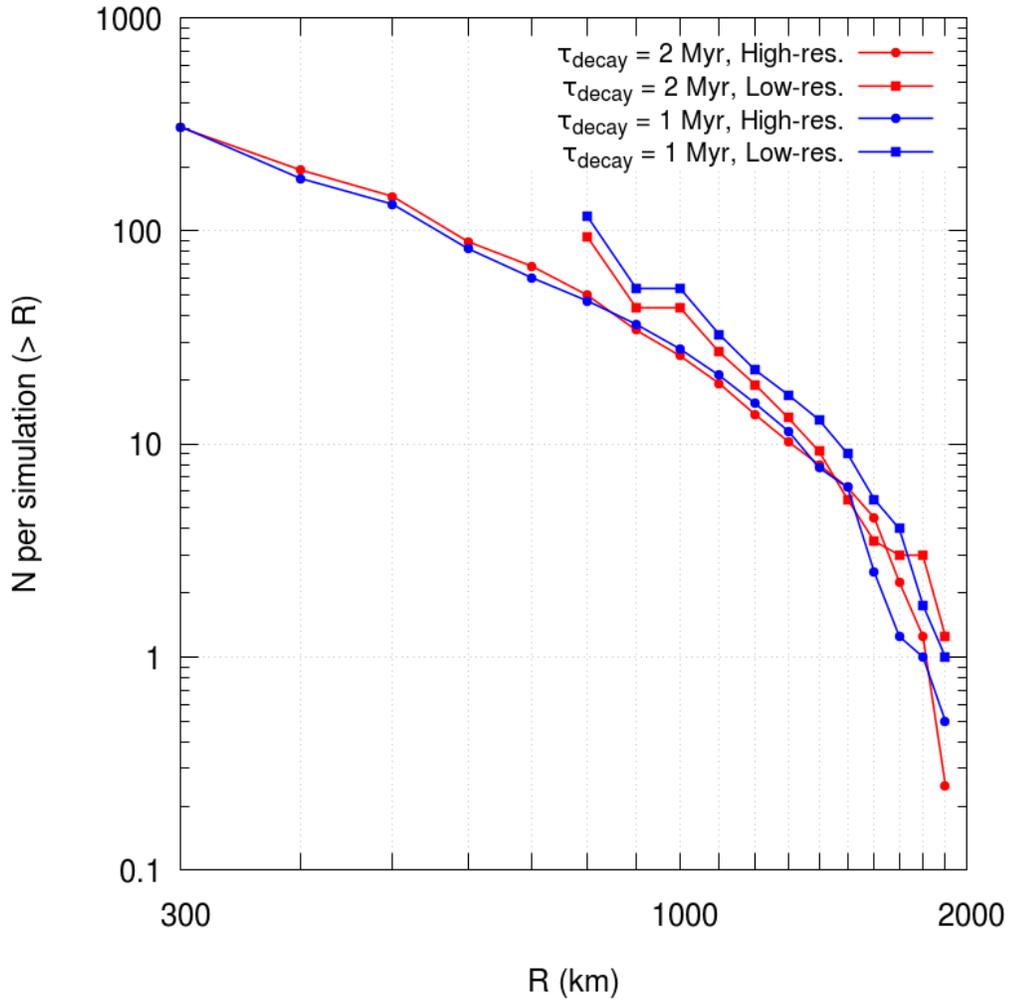

Fig. 6 - Size-frequency distribution (SFD) of objects with radii ($R$) less than 2000 km in EJS simulations (i.e. object with size smaller or similar to Moon) at 50 Myr. The red line represents data with gas dissipation timescale $\tau_{\text{decay}} = 2$ Myr and the blue line represents data with gas dissipation timescale $\tau_{\text{decay}} = 1$ Myr. The circles and squares represent high-resolution and low-resolution data, respectively. Note that the SFD become steeper for object with $R > 1000$ km, suggesting most of the mass of the leftover planetesimals are in smaller size objects ($R < 1000$ km).

Table 2 - Percentage of mass for objects with radii < 2000 km in different size bin at 50 Myr.

| | | R < 500 km | 500 km < R < 1000 km | 1000 km < R < 1500 km | 1500 km < R < 2000 km | Total mass ($M_{\text{Earth}}$) |
|---|---|---|---|---|---|---|
| | | | | | | |
| High-resolution | EJS, $\tau_{\mathrm{decay}} = 2$ Myr | 7.93 % | 35.0% | 30.6 % | 26.5 % | $0.24 \pm 0.03$ |
| | EJS, $\tau_{\mathrm{decay}} = 1$ Myr | 8.57 % | 31.3% | 35.0 % | 25.1 % | $0.22 \pm 0.02$ |
| Low-resolution | EJS, $\tau_{\mathrm{decay}} = 2$ Myr | NIL | 21.7 % | 52.4 % | 25.9 % | $0.20 \pm 0.03$ |
| | EJS, $\tau_{\mathrm{decay}} = 1$ Myr | NIL | 21.7 | 47.9 % | 30.4 % | $0.25 \pm 0.04$ |

## 4. Conclusions

We performed 30 sets of N-body simulations with the state-of-the-art GPU code *GENGA* (Grimm and Stadel, 2014) on top-of-the-line GPUs. These simulations began with only planetesimals in order to study the classical model of terrestrial planet formation, in which Jupiter and Saturn are placed close to their current orbits without migration. We considered two scenarios, where Jupiter and Saturn are either on their current eccentric orbits (EJS case; Chambers, 2001) or pre-instability circular orbits (CJS case; Tsiganis et al., 2005). In assessing these two scenarios, we focused on the dynamical configuration of the final system, as well as the timing of giant impacts on Earth analogues and the distribution of the leftover planetesimals during Earth's late accretion.

### 4.1. EJS is dynamically feasible, but could be cosmochemically problematic

Consistent with previous studies, we found that the EJS scenario is more successful in explaining the current dynamical features of the terrestrial planets, including planets with ~1 $M_{\mathrm{Earth}}$ in the current Venus-Earth region and a low mass of Mars with less than $0.3\ M_{\mathrm{Earth}}$. Assuming a shorter gas disc dissipation timescale ($\tau_{\mathrm{decay}} = 1$ Myr) would increase the chance of forming massive Venus-Earth pair. In the perspective of matching the chronology of the formation of Earth and Mars, our EJS simulations typically form Earth in ~50 to 80 Myr (Kleine et al., 2009; Rudge et al., 2010; Yin et al., 2002; cf. Lammer et al., 2020; Schiller et al., 2018) and Mars within 10 Myr (Dauphas and Pourmand, 2011; Tang and Dauphas, 2014).

However, a potential problem of the EJS model is the severe radial mixing caused by the sweeping of the $\nu_5$ secular resonance (Bromley and Kenyon, 2017; Nagasawa et al., 2000, 2005; Thommes et al., 2008). In W21b we have shown that the enhanced mixing would likely cause Earth-region embryos and Mars-region embryos to have similar feeding zones and hence similar isotopic composition. This could potentially violate the distinct isotopic compositions between them as inferred from isotopic measurements (e.g. Dauphas et al., 2014; Franchi et al., 1999; Trinquier et al., 2009, 2007) and mixing models (Brasser et al., 2018; Dauphas, 2017; Sanloup et al., 1999). Moreover, the potential existence of a spatial isotopic gradient in the inner solar system, which is inferred from the [54]Cr and [50]Ti measurements of the inner solar system meteorites (Mezger et al., 2020; Trinquier et al., 2009; Yamakawa et al., 2010), could also argue against the strong mixing effect during the formation of terrestrial planets. To further confirm these arguments, we will study the isotopic composition of Earth and Mars computed from our N-body simulations in detail in a future paper.

### 4.2. An early giant planet instability could be the solution of the CJS scenario

Either the current Earth-Mars isotopic differences are a low probability result in the EJS scenario, or the giant planets were residing in more circular orbits (CJS) during gas disc dissipation, as suggested by the Nice model (Tsiganis et al., 2005), in order to limit the mixing effect. However, we find that without the influence of the giant planet instability in the first 80 Myr, the CJS scenario is inconsistent with the current architecture of the inner solar system because it forms planets in the region of Mars that are in general more massive than in the region of Venus and Earth. To overcome this, it has been suggested that the giant planets should undergo a dynamical instability much earlier than previously thought, potentially within 80 Myr after gas disc dissipation (Clement et al., 2018; Costa et al., 2020; Mojzsis et al., 2019; Nesvorný et al., 2021; Ribeiro et al., 2020), so as to limit the growth of Mars (Clement et al., 2018; Nesvorný et al., 2021) and to trigger giant impacts to form massive planets in the Venus-Earth region (Clement et al., 2020). However, whether the early giant planet instability may induce too much radial mixing within the terrestrial planet region requires further investigation.

### 4.3. The last giant impact (GI) on Earth likely occurred before 80 Myr

Since we can now form Earth directly from a disc of small planetesimals, the collisional growth history of Earth can be revisited in our high-resolution N-body simulations. Focusing on the dynamically successful EJS simulations, we find that Earth analogues typically finish accreting over 90% of their mass before 80 Myr of the simulation. The Earth analogues also suffer more than 90% of GIs within 80 Myr, in which more than 70% of the Earth analogues suffer a final GI within 80 Myr. Our simulations therefore indicate a higher probability that the Moon-forming GI occurred early, within 80 Myr. If the Moon-forming GI is in fact the second-to-last GI instead of the final GI (Brasser et al., 2016a; Genda et al., 2017a), then its impact scale in general match better with the suggested scale of Moon-forming impact and its timing can be as early as ~20 to 30 Myr after the birth of the solar system. However, geochronological evidence for such an early impact is thin. An early Moon-forming GI around 50 Myr to 60 Myr better matches with the lunar crustal and core formation age (Barboni et al., 2017; Thiemens et al., 2019). However, such early Moon formation would subsequently require early strong tidal heating to explain Moon's long-lived magma ocean (Elkins-Tanton et al., 2011), which is why some authors have suggested a later formation age near 100 Myr based on Sm-Nd differentiation ages (e.g. Borg et al., 2011; Carlson et al., 2014; Gaffney and Borg, 2014; Maurice et al., 2020).

### 4.4. Late accretion could be delivered by both large and small impactors

Our high-resolution simulations are also fine-grained enough to allow us to study the size-frequency distribution (SFD) of the leftover planetesimals right after the assembly of terrestrial planets. The shape of the SFD has strong implications to the nature of late accretion (Chou, 1978) and its ability to deliver the excess in highly siderophile elements to the terrestrial, martian and lunar mantles. Although our simulations form about 10 Moneta-sized (~ 1 $M_{Moon}$; 1500 km < $R$ < 2000 km) objects at 50 Myr (roughly corresponding to the time after the second-to-last GI but before the final GI based on our simulation results), the slope of the SFD of the leftover planetesimals steepens after $R$ > 1000 km. We find that that most of the mass (~60 %) of the leftover planetesimals are in sub-Moneta sized to Ceres-sized objects (500 km < $R$ < 1500 km), and that late accretion to the terrestrial planets could be contributed by both large (i.e. Moneta-sized) and small ($R$ ~ $10^2$ km) impactors. The precise ratios of mass accreted to the terrestrial planets need to be determined from future studies.

### 4.5. Models with a lower initial mass in the asteroid belt merit further study

Our simulations yield ~0.25 $M_{Earth}$ of leftover planetesimals at 50 Myr that can contribute to late accretion. However, this amount of leftover planetesimals would likely result in an overshoot of HSE abundances in the terrestrial, martian and especially the lunar mantle (Brasser et al., 2016a, 2021; Day et al., 2007, 2016; Day and Walker, 2015; Raymond et al., 2013; Walker, 2009). One potential solution is to begin with a lower mass or even empty initial asteroid belt (Izidoro et al., 2014, 2015; Mah and Brasser, 2021; Raymond and Izidoro, 2017) than the MMSN, so that the sweeping secular resonance would not implant too much mass (< 0.5 $M_{Earth}$) into the terrestrial planet crossing orbits. This can also slightly mitigate the strong radial mixing effect in the EJS simulations (W21a). However, an entirely empty asteroid belt, as suggested by Raymond and Izidoro (2017) is difficult to reconcile with the various isotopic compositions of different non-carbonaceous meteorites (Mezger et al., 2020; Trinquier et al., 2009; Yamakawa et al., 2010), and we therefore advocate that there was initially some low amount of mass (< $10^{-1}$ $M_{Earth}$) in the asteroid belt region.

While the study presented here represents a significant improvement over previous studies in terms of mass resolution, we reiterate that our N-body simulations still assume that all collisions result in perfect merging. This is an important caveat to note because recent hydrodynamical studies of pairwise collisions between rocky bodies (e.g. Genda et al., 2017b; Leinhardt and Stewart, 2012; Timpe et al., 2020) have shown that the outcomes of most collisions are disruptive rather than accretionary and only rarely result in perfect merging. Imperfect collisions resulting in fragmentation have been shown to be crucial in determining the growth timescale of embryos and planets (Kobayashi and Dauphas, 2013; Quintana et al., 2016). Hence, future studies should include collisional fragmentation and use increasingly high-resolution N-body simulations in order to more realistically deduce the timing and consequences of GIs, as well as the SFD of the leftover planetesimals in the terrestrial planet and the asteroid belt region.

### Acknowledgments

This work has been carried out within the framework of the National Center of Competence in Research PlanetS, supported by the Swiss National Science Foundation (SNSF). The authors acknowledge the financial support of the SNSF. The authors acknowledge the computational support from Service and Support for Science IT (S³IT) of University of Zurich and the Swiss National Supercomputing Centre (CSCS).

### Appendix. Statistics for the orbital architecture of the system

Here we present a summary of the dynamical results from our GPU N-body simulations by showing the overall mass-distance distribution of the planets, as well as the orbital statistics that describe the dynamical features of each set of simulations. Fig. 7 demonstrates the mass-distance distribution of the planets from all simulations performed. One of the most obvious features is that in the CJS simulations, we do not form a single planet with mass close to ~1 $M_{Earth}$ after 80 Myr in the current Venus-Earth region. Furthermore, planets formed in the Mars region are in general too massive compared to those in Venus and Earth. This denotes the dynamical failure of the CJS case. We presented an example of high-resolution simulations that represent the typical CJS results in Fig. 2. Hence, our results, together with other previous studies (Clement et al., 2018; Costa et al., 2020; Mojzsis et al., 2019; Ribeiro et al., 2020), argue for a giant planet instability occurring within at least 80 Myr of the solar system formation, so that the orbits of Mars can be cleared before an over massive Mars analogue forms in that region (Clement et al., 2018).

On the other hand, the EJS simulations are more successful in forming planets close to an Earth mass at ~ 1 AU and a less massive Mars within 150 Myr, although the majority of Mars analogues are still with mass > 0.3 $M_{Mars}$. Comparing the $\tau_{decay} = 1$ and 2 Myr of the EJS results, we found that shorter-lived discs form more massive Earth and Venus analogues. There are four planets across 9 simulations with mass greater than 1 $M_{Earth}$ when $\tau_{decay} = 1$ Myr, but only one of them in 9 simulations of $\tau_{decay} = 2$ Myr. This is because the mass loss of embryos through Type-I migration (Tanaka et al., 2002; Tanaka and Ward, 2004) to the Sun is larger when the gas disc lifetime is longer. While successfully forming an Earth-like planet at 1 AU, all EJS simulations tend to form planets that are too massive compared to Mercury at ~0.4 AU. Explaining the formation of Mercury should rely on special collision treatments, for example repeated hit and run collisions of originally larger planets (Asphaug et al., 2006; Chau et al., 2018; Svetsov, 2011), or fragmenting collisions between two large protoplanets (Asphaug, 2010; Asphaug and Reufer, 2014; Benz et al., 2007, 1988; Chau et al., 2018), which are procedures not accounted for in our simulations. A typical example of the high-resolution EJS simulation is shown in Fig. 1.

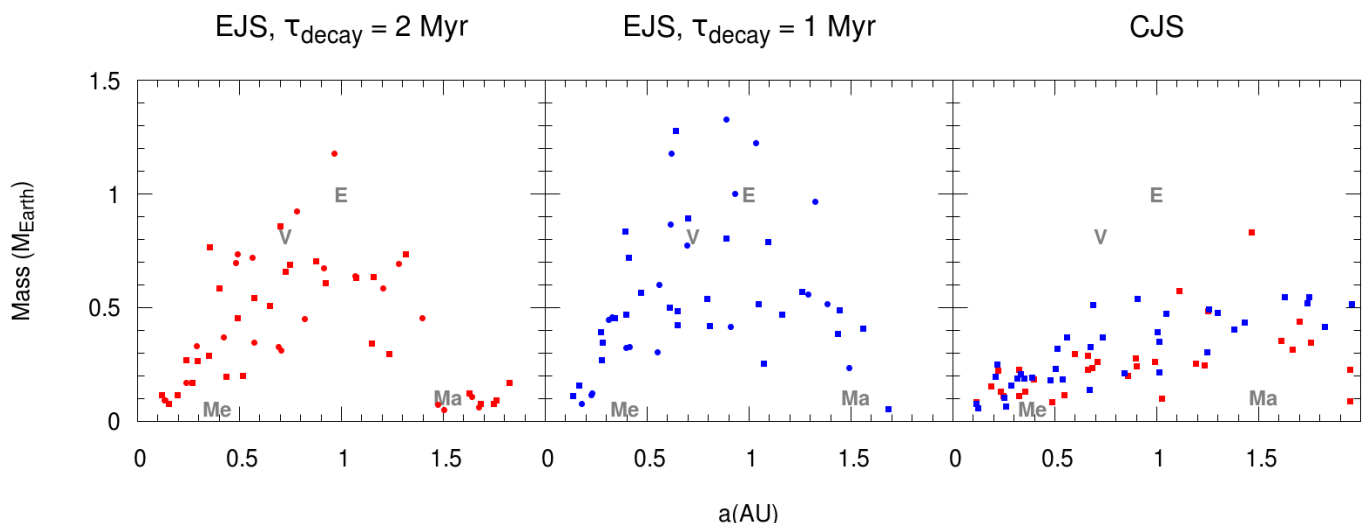

Fig. 7 - Mass-distance distribution of all the planets at the end of simulations at 150 Myr for EJS and CJS. Circles represent data from the high-resolution simulations ($r = 350$ km) and squares represent data from low-resolution simulations ($r = 800$ km). Data from $\tau_{decay} = 2$ Myr are in red and those from $\tau_{decay} = 1$ Myr

are in blue. The grey capital letters represent the current mass of the terrestrial planets. In general, EJS is more successful in reproducing the current orbital architecture of the terrestrial planets.

We present also the orbital statistics describing the orbital architecture of the system. These orbital statistics are adapted from Chambers (2001). Table 3 shows the average statistical properties of different sets of simulations. $N_{pl}$ is the number of planets at 150 Myr for EJS and CJS in the terrestrial planet region ($a < 2$ AU). $S_c$ is the concentration parameter which measures the degree of mass concentration in one part of the system, given by

$$S_c = \max \left( \frac{\sum_k \mu_k}{\sum_k \mu_k (\log(a/a_k))^2} \right),$$ (1)

where $\mu_k = m_k / M_\odot$. AMD is the angular momentum deficit which measure how the degree of dynamical excitation of the orbits, given by

$$\text{AMD} = \frac{\sum_k \mu_k \sqrt{a_k} [1 - \sqrt{(1 - e_k^2)} \cos I_k]}{\sum_k \mu_k \sqrt{a_k}}.$$ (2)

Finally, $S_H$ is the mean orbital spacing between planets scaled by mutual Hill radii, which is

$$S_H = \frac{2}{N-1} \sum_{k=1}^{N-1} \frac{a_{k+1} - a_k}{a_{k+1} + a_k} \left( \frac{\mu_{k+1} + \mu_k}{3} \right)^{-1/3}.$$ (3)

According to Table 3, all the simulation sets tend to yield a system which has higher $N_{pl}$ and AMD, but lower $S_C$ than the current system. The mean spacing $S_H$ is comparable to the current system for all sets of simulation. There is still a chance of lowering $N_{pl}$ through collision between planets, especially in the Mercury region where the planets are generally more closely spaced due to their smaller Hill sphere (see also Fig. 1). This late giant impact on proto-Mercury could provide an explanation on its current small size and high core to mantle mass ratio (Asphaug et al., 2006; Benz et al., 1988; Chau et al., 2018; Svetsov, 2011). The lower $S_C$ than the current system can be explained by forming massive planets in the region of Mercury in EJS. In some cases the Mars region planets can be as massive as ~4 to 5 $M_{Mars}$, which also lowers the $S_C$ of the computed planetary systems. Although the AMD of the EJS simulations is higher than the current system, it is still within $1\sigma$ of the actual mass and given the high mass in leftover planetesimals ($>10^{-2} M_{Earth}$ at 150 Myr) in our simulations it may become even lower still. There is a dependence of the final AMD on the initial number of planetesimals. Our low-resolution simulations tend to have a higher AMD because there are fewer particles in the system causing a weaker dynamical friction experienced by the planets even though the total mass is about the same.

We only present results for the CJS simulations with the low-resolution because we only run the low-resolution CJS simulations to 150 Myr. Compared to the low-resolution results of the EJS simulations, the CJS cases tend to form a system with a higher number of $N_{pl}$, an even lower $S_C$ and a lower AMD. Due to the negligible effect of the sweeping secular resonance, planetary systems are less excited in the CJS simulation. The lower dynamical excitation of the systems also explains the high number of planets formed as the lower AMD cases in general suffer fewer embryo merging collisions. The lower $S_C$ of the CJS simulations are due to many massive planets forming in the Mars region, causing the system's mass to be more evenly distributed throughout the terrestrial planet region.

To summarize, the configuration of parameters that dynamically matches the current system most closely is the high-resolution EJS case with $\tau_{decay} = 1$ Myr. However, we have to emphasise that the statistics are based only on a very limited number of simulations (we performed only 8 sets of high-resolution simulations in total for EJS). Hence, more high-resolution simulations are required to ensure that the statistics presented in Table 3 are robust.

Table 3 - Properties of the terrestrial planet system at the end of simulation. $N_{pl}$ is the average number of planets, $S_C$ is the average concentration parameter (Eq. 1), AMD is the average angular momentum deficit of the planets normalised to the current system AMD (Eq. 2) and $S_H$ is the average orbital spacing of the planets (Eq. 3). The uncertainties are $1\sigma$. Note that we do not show results for the high-resolution CJS results because they are only performed to 80 Myr.

| | | Number of simulations | $N_{pl}$ | $S_C$ | AMD | $S_H$ |
|---|---|---|---|---|---|---|
| EJS, $\tau_{decay} = 2$ Myr | High-resolution | 4 | $5.75 \pm 1.48$ | $22.4 \pm 0.82$ | $1.73 \pm 1.43$ | $44.2 \pm 5.87$ |
| | Low-resolution | 5 | $5.80 \pm 0.98$ | $17.8 \pm 2.20$ | $3.87 \pm 2.80$ | $48.9 \pm 8.61$ |
| EJS, $\tau_{decay} = 1$ Myr | High-resolution | 4 | $5.00 \pm 0.71$ | $24.9 \pm 2.94$ | $1.46 \pm 0.65$ | $42.3 \pm 1.89$ |
| | Low-resolution | 5 | $5.40 \pm 1.50$ | $20.6 \pm 6.82$ | $2.47 \pm 0.74$ | $41.9 \pm 6.52$ |
| CJS, $\tau_{decay} = 1$ or 2 Myr | Low-resolution | 10 | $6.80 \pm 0.98$ | $13.6 \pm 3.20$ | $1.69 \pm 1.18$ | $47.7 \pm 9.09$ |
| Current system | | | 4 | 89.7 | 1 | 43.2 |

The sweeping secular resonances is the most decisive effect in the EJS simulations in terms of shaping the dynamical features of the final system, as well as altering the composition of the embryos and possibly the final planets (W21a,b). The sweeping secular resonances could also be the cause of the depletion of mass in the asteroid belt (e.g. Bromley and Kenyon, 2017; Nagasawa et al., 2000, 2005; Thommes et al., 2008), since the asteroid belt material is removed in this process. Table 4 records the mass left in the asteroid belt at the end of each high-resolution simulation. More than 99.5% of the original mass is removed from the asteroid belt. The removal efficiency is higher when the gas disc lifetime is longer, because the $\nu_5$ resonance sweeps through the asteroid belt at a slower rate. Apparently, there is too little mass left behind in the asteroid belt when $\tau_{decay} = 2$ Myr, but too much mass when $\tau_{decay} = 1$ Myr. The initial mass of a planetesimal in our high-resolution simulation is ~$9 \times 10^{-5}$ $M_{Earth}$, which is about one-fourth of the current mass of the asteroid belt. Hence, higher resolution simulations, together with fragmentation implementation into the N-body code, is required to further study the effect of the sweeping secular resonance on depleting the asteroid belt.

Table 4 - Remaining mass in the asteroid belt (2 AU < $a$ < 3.2 AU) on non-Mars crossing orbits in each high-resolution simulation (initial planetesimals' $r$ = 350 km). $M_{ast,f}$ is the final mass of the asteroid belt and $M_{ast,i}$ = 1.05 $M_{Earth}$ is the initial mass in the asteroid belt.

| | Remnant mass of the asteroid belt ($M_{ast,f}$ in $M_{Earth}$) | | | | Average $M_{ast,f}/M_{ast,i}$ |
|---|---|---|---|---|---|
| | run1 | run2 | run3 | run4 | |
| EJS, $\tau_{decay}$ = 2 Myr | 0 | $9.02 \times 10^{-5}$ | 0 | $7.22 \times 10^{-4}$ | $1.93 \times 10^{-4}$ |
| EJS, $\tau_{decay}$ = 1 Myr | $3.52 \times 10^{-3}$ | $2.89 \times 10^{-3}$ | $4.60 \times 10^{-3}$ | $2.16 \times 10^{-3}$ | $3.13 \times 10^{-3}$ |

## Reference

Albarede, F., Martine, J., 1984. Unscrambling the lead model ages. Geochim. Cosmochim. Acta 48, 207–212. https://doi.org/10.1016/0016-7037(84)90364-8

Allègre, C.J., Manhès, G., Göpel, C., 2008. The major differentiation of the Earth at ~4.45 Ga. Earth Planet. Sci. Lett. 267, 386–398. https://doi.org/10.1016/j.epsl.2007.11.056

Asphaug, E., 2010. Similar-sized collisions and the diversity of planets. Geochemistry 70, 199–219. https://doi.org/10.1016/j.chemer.2010.01.004

Asphaug, E., Agnor, C.B., Williams, Q., 2006. Hit-and-run planetary collisions. Nature 439, 155–160. https://doi.org/10.1038/nature04311

Asphaug, E., Reufer, A., 2014. Mercury and other iron-rich planetary bodies as relics of inefficient accretion. Nat. Geosci. 7, 564–568. https://doi.org/10.1038/ngeo2189

Avice, G., Marty, B., 2014. The iodine–plutonium–xenon age of the Moon–Earth system revisited. Philos. Trans. R. Soc. Math. Phys. Eng. Sci. 372, 20130260. https://doi.org/10.1098/rsta.2013.0260

Barboni, M., Boehnke, P., Keller, B., Kohl, I.E., Schoene, B., Young, E.D., McKeegan, K.D., 2017. Early formation of the Moon 4.51 billion years ago. Sci. Adv. 3, e1602365. https://doi.org/10.1126/sciadv.1602365

Benz, W., Anic, A., Horner, J., Whitby, J.A., 2007. The Origin of Mercury. Space Sci. Rev. 132, 189–202. https://doi.org/10.1007/s11214-007-9284-1

Benz, W., Slattery, W.L., Cameron, A.G.W., 1988. Collisional stripping of Mercury's mantle. Icarus 74, 516–528. https://doi.org/10.1016/0019-1035(88)90118-2

Benz, W., Slattery, W.L., Cameron, A.G.W., 1986. The origin of the moon and the single-impact hypothesis I. Icarus 66, 515–535. https://doi.org/10.1016/0019-1035(86)90088-6

Borg, L.E., Cassata, W.S., Wimpenny, J., Gaffney, A.M., Shearer, C.K., 2020. The formation and evolution of the Moon's crust inferred from the Sm-Nd isotopic systematics of highlands rocks. Geochim. Cosmochim. Acta 290, 312–332. https://doi.org/10.1016/j.gca.2020.09.013

Borg, L.E., Connelly, J.N., Boyet, M., Carlson, R.W., 2011. Chronological evidence that the Moon is either young or did not have a global magma ocean. Nature 477, 70–72. https://doi.org/10.1038/nature10328

Borg, L.E., Gaffney, A.M., Kruijer, T.S., Marks, N.A., Sio, C.K., Wimpenny, J., 2019. Isotopic evidence for a young lunar magma ocean. Earth Planet. Sci. Lett. 523, 115706. https://doi.org/10.1016/j.epsl.2019.07.008

Borg, L.E., Gaffney, A.M., Shearer, C.K., 2015. A review of lunar chronology revealing a preponderance of 4.34–4.37 Ga ages. Meteorit. Planet. Sci. 50, 715–732. https://doi.org/10.1111/maps.12373

Boss, A.P., 1997. Giant Planet Formation by Gravitational Instability. Science 276, 1836–1839. https://doi.org/10.1126/science.276.5320.1836

Bottke, W.F., Andrews-Hanna, J.C., 2017. A post-accretionary lull in large impacts on early Mars. Nat. Geosci. 10, 344–348. https://doi.org/10.1038/ngeo2937

Bottke, W.F., Walker, R.J., Day, J.M.D., Nesvorny, D., Elkins-Tanton, L., 2010. Stochastic Late Accretion to Earth, the Moon, and Mars. Science 330, 1527–1530. https://doi.org/10.1126/science.1196874

Bouvier, A., Wadhwa, M., 2010. The age of the solar system redefined by the oldest Pb–Pb age of a meteoritic inclusion. Nat. Geosci. 3, 637–641. https://doi.org/10.1038/ngeo941

Bouvier, L.C., Costa, M.M., Connelly, J.N., Jensen, N.K., Wielandt, D., Storey, M., Nemchin, A.A., Whitehouse, M.J., Snape, J.F., Bellucci, J.J., Moynier, F., Agranier, A., Gueguen, B., Schönbächler, M., Bizzarro, M., 2018. Evidence for extremely rapid magma ocean crystallization and crust formation on Mars. Nature 558, 586–589. https://doi.org/10.1038/s41586-018-0222-z

Boyet, M., Carlson, R.W., Borg, L.E., Horan, M., 2015. Sm–Nd systematics of lunar ferroan anorthositic suite rocks: Constraints on lunar crust formation. Geochim. Cosmochim. Acta 148, 203–218. https://doi.org/10.1016/j.gca.2014.09.021

Brasser, R., Dauphas, N., Mojzsis, S.J., 2018. Jupiter's Influence on the Building Blocks of Mars and Earth. Geophys. Res. Lett. 45, 5908–5917. https://doi.org/10.1029/2018GL078011

Brasser, R., Matsumura, S., Ida, S., Mojzsis, S.J., Werner, S.C., 2016a. ANALYSIS OF TERRESTRIAL PLANET FORMATION BY THE GRAND TACK MODEL: SYSTEM ARCHITECTURE AND TACK LOCATION. Astrophys. J. 821, 75. https://doi.org/10.3847/0004-637X/821/2/75

Brasser, R., Mojzsis, S.J., Matsumura, S., Ida, S., 2017. The cool and distant formation of Mars. Earth Planet. Sci. Lett. 468, 85–93. https://doi.org/10.1016/j.epsl.2017.04.005

Brasser, R., Mojzsis, S.J., Werner, S.C., Abramov, O., 2021. A new estimate for the age of highly-siderophile element retention in the lunar mantle from late accretion. Icarus 361, 114389. https://doi.org/10.1016/j.icarus.2021.114389

Brasser, R., Mojzsis, S.J., Werner, S.C., Matsumura, S., Ida, S., 2016b. Late veneer and late accretion to the terrestrial planets. Earth Planet. Sci. Lett. 455, 85–93. https://doi.org/10.1016/j.epsl.2016.09.013

Brasser, R., Werner, S.C., Mojzsis, S.J., 2020. Impact bombardment chronology of the terrestrial planets from 4.5 Ga to 3.5 Ga. Icarus 338, 113514. https://doi.org/10.1016/j.icarus.2019.113514

Bromley, B.C., Kenyon, S.J., 2017. Terrestrial Planet Formation: Dynamical Shake-up and the Low Mass of Mars. Astron. J. 153, 216. https://doi.org/10.3847/1538-3881/aa6aaa

Cameron, A.G.W., Ward, W.R., 1976. The Origin of the Moon 7.

Canup, R., Salmon, J., 2018. Origin of Phobos and Deimos by the impact of a Vesta-to-Ceres sized body with Mars. Sci. Adv. 4, eaar6887. https://doi.org/10.1126/sciadv.aar6887

Canup, R.M., Asphaug, E., 2001. Origin of the Moon in a giant impact near the end of the Earth's formation. Nature 412, 708–712. https://doi.org/10.1038/35089010

Canup, R.M., 2012. Forming a Moon with an Earth-like Composition via a Giant Impact. Science, 338, 1052-1055

Caracausi, A., Avice, G., Burnard, P.G., Füri, E., Marty, B., 2016. Chondritic xenon in the Earth's mantle. Nature 533, 82–85. https://doi.org/10.1038/nature17434

Carlson, R.W., Borg, L.E., Gaffney, A.M., Boyet, M., 2014. Rb-Sr, Sm-Nd and Lu-Hf isotope systematics of the lunar Mg-suite: the age of the lunar crust and its relation to the time of Moon formation. Philos. Trans. R. Soc. Math. Phys. Eng. Sci. 372, 20130246. https://doi.org/10.1098/rsta.2013.0246

Carter, P.J., Leinhardt, Z.M., Elliott, T., Walter, M.J., Stewart, S.T., 2015. COMPOSITIONAL EVOLUTION DURING ROCKY PROTOPLANET ACCRETION. Astrophys. J. 813, 72. https://doi.org/10.1088/0004-637X/813/1/72

Chambers, J.E., 2001. Making More Terrestrial Planets. Icarus 152, 205–224. https://doi.org/10.1006/icar.2001.6639

Chambers, J.E., 1999. A hybrid symplectic integrator that permits close encounters between massive bodies. Mon. Not. R. Astron. Soc. 304, 793–799. https://doi.org/10.1046/j.1365-8711.1999.02379.x

Chau, A., Reinhardt, C., Helled, R., Stadel, J., 2018. Forming Mercury by Giant Impacts. Astrophys. J. 865, 35. https://doi.org/10.3847/1538-4357/aad8b0

Chou, C.-L., 1978. Fractionation of Siderophile Elements in the Earth's Upper Mantle. Proc. 9th Lunar Planet. Sci. Conf. Houst. TX 219–230.

Citron, R.I., Genda, H., Ida, S., 2015. Formation of Phobos and Deimos via a giant impact. Icarus 252, 334–338. https://doi.org/10.1016/j.icarus.2015.02.011

Clement, M.S., Chambers, J.E., 2021. Dynamical avenues for Mercury's origin II: in-situ formation in the inner terrestrial disk. ArXiv210411252 Astro-Ph.

Clement, M.S., Chambers, J.E., Jackson, A.P., 2021a. Dynamical Avenues for Mercury's Origin. I. The Lone Primordial Survivor of a Primordial Generation of Short-period Protoplanets. Astron. J. 161, 240. https://doi.org/10.3847/1538-3881/abf09f

Clement, M.S., Kaib, N.A., Chambers, J.E., 2020. Embryo Formation with GPU Acceleration: Reevaluating the Initial Conditions for Terrestrial Accretion. Planet. Sci. J. 1, 18. https://doi.org/10.3847/PSJ/ab91aa

Clement, M.S., Kaib, N.A., Chambers, J.E., 2019. Dynamical Constraints on Mercury's Collisional Origin. Astron. J. 157, 208. https://doi.org/10.3847/1538-3881/ab164f

Clement, M.S., Kaib, N.A., Raymond, S.N., Chambers, J.E., 2021b. The early instability scenario: Mars' mass explained by Jupiter's orbit. Icarus 367, 114585. https://doi.org/10.1016/j.icarus.2021.114585

Clement, M.S., Kaib, N.A., Raymond, S.N., Walsh, K.J., 2018. Mars' growth stunted by an early giant planet instability. Icarus 311, 340–356. https://doi.org/10.1016/j.icarus.2018.04.008

Clement, M.S., Raymond, S.N., Kaib, N.A., Deienno, R., Chambers, J.E., Izidoro, A., 2021c. Born eccentric: Constraints on Jupiter and Saturn's pre-instability orbits. Icarus 355, 114122. https://doi.org/10.1016/j.icarus.2020.114122

Costa, M.M., Jensen, N.K., Bouvier, L.C., Connelly, J.N., Mikouchi, T., Horstwood, M.S.A., Suuronen, J.-P., Moynier, F., Deng, Z., Agranier, A., Martin, L.A.J., Johnson, T.E., Nemchin, A.A., Bizzarro, M., 2020. The internal structure and geodynamics of Mars inferred from a 4.2-Gyr zircon record. Proc. Natl. Acad. Sci. 117, 30973–30979. https://doi.org/10.1073/pnas.2016326117

Craddock, R.A., 2011. Are Phobos and Deimos the result of a giant impact? Icarus 211, 1150–1161. https://doi.org/10.1016/j.icarus.2010.10.023

Dauphas, N., 2017. The isotopic nature of the Earth's accreting material through time. Nature 541, 521–524. https://doi.org/10.1038/nature20830

Dauphas, N., Chen, J.H., Zhang, J., Papanastassiou, D.A., Davis, A.M., Travaglio, C., 2014. Calcium-48 isotopic anomalies in bulk chondrites and achondrites: Evidence for a uniform isotopic reservoir in the inner protoplanetary disk. Earth Planet. Sci. Lett. 407, 96–108. https://doi.org/10.1016/j.epsl.2014.09.015

Dauphas, N., Pourmand, A., 2011. Hf–W–Th evidence for rapid growth of Mars and its status as a planetary embryo. Nature 473, 489–492. https://doi.org/10.1038/nature10077

Day, J.M.D., Brandon, A.D., Walker, R.J., 2016. Highly Siderophile Elements in Earth, Mars, the Moon, and Asteroids. Rev. Mineral. Geochem. 81, 161–238. https://doi.org/10.2138/rmg.2016.81.04

Day, J.M.D., Pearson, D.G., Taylor, L.A., 2007. Highly Siderophile Element Constraints on Accretion and Differentiation of the Earth-Moon System. Science 315, 217–219. https://doi.org/10.1126/science.1133355

Day, J.M.D., Walker, R.J., 2015. Highly siderophile element depletion in the Moon. Earth Planet. Sci. Lett. 423, 114–124. https://doi.org/10.1016/j.epsl.2015.05.001

Duncan, M.J., Levison, H.F., Lee, M.H., 1998. A Multiple Time Step Symplectic Algorithm for

Integrating Close Encounters. Astron. J. 116, 2067. https://doi.org/10.1086/300541

Elkins-Tanton, L.T., Burgess, S., Yin, Q.-Z., 2011. The lunar magma ocean: Reconciling the solidification process with lunar petrology and geochronology. Earth Planet. Sci. Lett. 304, 326–336. https://doi.org/10.1016/j.epsl.2011.02.004

Franchi, I.A., Wright, I.P., Sexton, A.S., Pillinger, C.T., 1999. The oxygen-isotopic composition of Earth and Mars. Meteorit. Planet. Sci. 34, 657–661. https://doi.org/10.1111/j.1945-5100.1999.tb01371.x

Gaffney, A.M., Borg, L.E., 2014. A young solidification age for the lunar magma ocean. Geochim. Cosmochim. Acta 140, 227–240. https://doi.org/10.1016/j.gca.2014.05.028

Genda, H., Brasser, R., Mojzsis, S.J., 2017a. The terrestrial late veneer from core disruption of a lunar-sized impactor. Earth Planet. Sci. Lett. 480, 25–32. https://doi.org/10.1016/j.epsl.2017.09.041

Genda, H., Iizuka, T., Sasaki, T., Ueno, Y., Ikoma, M., 2017b. Ejection of iron-bearing giant-impact fragments and the dynamical and geochemical influence of the fragment re-accretion. Earth Planet. Sci. Lett. 470, 87–95. https://doi.org/10.1016/j.epsl.2017.04.035

Gomes, R., Levison, H.F., Tsiganis, K., Morbidelli, A., 2005. Origin of the cataclysmic Late Heavy Bombardment period of the terrestrial planets. Nature 435, 466–469. https://doi.org/10.1038/nature03676

Grimm, S.L., Stadel, J.G., 2014. THE GENGA CODE: GRAVITATIONAL ENCOUNTERS INN-BODY SIMULATIONS WITH GPU ACCELERATION. Astrophys. J. 796, 23. https://doi.org/10.1088/0004-637X/796/1/23

Halliday, A., Rehkämper, M., Lee, D.-C., Yi, W., 1996. Early evolution of the Earth and Moon: new constraints from Hf-W isotope geochemistry. Earth Planet. Sci. Lett. 142, 75–89. https://doi.org/10.1016/0012-821X(96)00096-9

Halliday, A.N., 2000. Terrestrial accretion rates and the origin of the Moon. Earth Planet. Sci. Lett. 176, 17–30. https://doi.org/10.1016/S0012-821X(99)00317-9

Hartmann, W.K., Davis, D.R., 1975. Satellite-sized planetesimals and lunar origin. Icarus 24, 504–515. https://doi.org/10.1016/0019-1035(75)90070-6

Hayashi, C., 1981. Structure of the Solar Nebula, Growth and Decay of Magnetic Fields and Effects of Magnetic and Turbulent Viscosities on the Nebula. Prog. Theor. Phys. Suppl. 70, 35–53. https://doi.org/10.1143/PTPS.70.35

Hoffmann, V., Grimm, S.L., Moore, B., Stadel, J., 2017. Stochasticity and predictability in terrestrial planet formation. Mon. Not. R. Astron. Soc. 465, 2170–2188. https://doi.org/10.1093/mnras/stw2856

Hopkins, M.D., Mojzsis, S.J., 2015. A protracted timeline for lunar bombardment from mineral chemistry, Ti thermometry and U–Pb geochronology of Apollo 14 melt breccia zircons. Contrib. Mineral. Petrol. 169, 30. https://doi.org/10.1007/s00410-015-1123-x

Hosono, N., Karato, S., Makino, J., Saitoh, T.R., 2019. Terrestrial magma ocean origin of the Moon. Nat. Geosci. 12, 418–423. https://doi.org/10.1038/s41561-019-0354-2

Izidoro, A., Haghighipour, N., Winter, O.C., Tsuchida, M., 2014. TERRESTRIAL PLANET FORMATION IN A PROTOPLANETARY DISK WITH A LOCAL MASS DEPLETION: A SUCCESSFUL SCENARIO FOR THE FORMATION OF MARS. Astrophys. J. 782, 31. https://doi.org/10.1088/0004-637X/782/1/31

Izidoro, A., Raymond, S.N., Morbidelli, A., Winter, O.C., 2015. Terrestrial planet formation constrained by Mars and the structure of the asteroid belt. Mon. Not. R. Astron. Soc. 453, 3619–3634. https://doi.org/10.1093/mnras/stv1835

Jacobson, S.A., Morbidelli, A., 2014. Lunar and terrestrial planet formation in the Grand Tack scenario. Philos. Trans. R. Soc. Math. Phys. Eng. Sci. 372, 20130174. https://doi.org/10.1098/rsta.2013.0174

Kaib, N.A., Cowan, N.B., 2015. The feeding zones of terrestrial planets and insights into Moon formation. Icarus 252, 161–174. https://doi.org/10.1016/j.icarus.2015.01.013

Kleine, T., Münker, C., Mezger, K., Palme, H., 2002. Rapid accretion and early core formation on asteroids and the terrestrial planets from Hf–W chronometry. Nature 418, 952–955. https://doi.org/10.1038/nature00982

Kleine, T., Touboul, M., Bourdon, B., Nimmo, F., Mezger, K., Palme, H., Jacobsen, S.B., Yin, Q.-Z., Halliday, A.N., 2009. Hf–W chronology of the accretion and early evolution of asteroids and terrestrial planets. Geochim. Cosmochim. Acta, The Chronology of Meteorites and the Early solar system 73, 5150–5188. https://doi.org/10.1016/j.gca.2008.11.047

Kobayashi, H., Dauphas, N., 2013. Small planetesimals in a massive disk formed Mars. Icarus 225, 122–130. https://doi.org/10.1016/j.icarus.2013.03.006

Kokubo, E., Ida, S., 2002. Formation of Protoplanet Systems and Diversity of Planetary Systems. Astrophys. J. 581, 666. https://doi.org/10.1086/344105

Kokubo, E., Ida, S., 2000. Formation of Protoplanets from Planetesimals in the Solar Nebula. Icarus 143, 15–27. https://doi.org/10.1006/icar.1999.6237

Kokubo, E., Ida, S., 1998. Oligarchic Growth of Protoplanets. Icarus 131, 171–178. https://doi.org/10.1006/icar.1997.5840

Kokubo, E., Ida, S., 1996. On Runaway Growth of Planetesimals. Icarus 123, 180–191. https://doi.org/10.1006/icar.1996.0148

Kruijer, T.S., Kleine, T., Fischer-Gödde, M., Sprung, P., 2015. Lunar tungsten isotopic evidence for the late veneer. Nature 520, 534–537. https://doi.org/10.1038/nature14360

Lammer, H., Leitzinger, M., Scherf, M., Odert, P., Burger, C., Kubyshkina, D., Johnstone, C., Maindl, T., Schäfer, C.M., Güdel, M., Tosi, N., Nikolaou, A., Marcq, E., Erkaev, N.V., Noack, L., Kislyakova, K.G., Fossati, L., Pilat-Lohinger, E., Ragossnig, F., Dorfi, E.A., 2020. Constraining the early evolution of Venus and Earth through atmospheric Ar, Ne isotope and bulk K/U ratios. Icarus 339, 113551. https://doi.org/10.1016/j.icarus.2019.113551

Leinhardt, Z.M., Stewart, S.T., 2012. COLLISIONS BETWEEN GRAVITY-DOMINATED BODIES. I. OUTCOME REGIMES AND SCALING LAWS. Astrophys. J. 745, 79. https://doi.org/10.1088/0004-637X/745/1/79

Levison, H.F., Kretke, K.A., Duncan, M.J., 2015. Growing the gas-giant planets by the gradual accumulation of pebbles. Nature 524, 322–324. https://doi.org/10.1038/nature14675

Lodders, K., Fegley, B., 1997. An Oxygen Isotope Model for the Composition of Mars. Icarus 126, 373–394. https://doi.org/10.1006/icar.1996.5653

Lykawka, P.S., Ito, T., 2019. Constraining the Formation of the Four Terrestrial Planets in the solar system. Astrophys. J. 883, 130. https://doi.org/10.3847/1538-4357/ab3b0a

Mah, J., Brasser, R., 2021. Isotopically distinct terrestrial planets via local accretion. Icarus 354, 114052. https://doi.org/10.1016/j.icarus.2020.114052

Manhes, G., Allègre, C.J., Dupré, B., Hamelin, B., 1979. Lead-lead systematics, the "age of the Earth" and the chemical evolution of our planet in a new representation space. Earth Planet. Sci. Lett. 44, 91–104. https://doi.org/10.1016/0012-821X(79)90011-6

Marchi, S., Walker, R.J., Canup, R.M., 2020. A compositionally heterogeneous martian mantle due to late accretion. Sci. Adv. 6, eaay2338. https://doi.org/10.1126/sciadv.aay2338

Marinova, M.M., Aharonson, O., Asphaug, E., 2008. Mega-impact formation of the Mars hemispheric dichotomy. Nature 453, 1216–1219. https://doi.org/10.1038/nature07070

Maurice, M., Tosi, N., Schwinger, S., Breuer, D., Kleine, T., 2020. A long-lived magma ocean on a young Moon. Sci. Adv. 6, eaba8949. https://doi.org/10.1126/sciadv.aba8949

Mezger, K., Schönbächler, M., Bouvier, A., 2020. Accretion of the Earth—Missing Components? Space Sci. Rev. 216, 27. https://doi.org/10.1007/s11214-020-00649-y

Mojzsis, S.J., Brasser, R., Kelly, N.M., Abramov, O., Werner, S.C., 2019. Onset of Giant Planet Migration before 4480 Million Years Ago. Astrophys. J. 881, 44. https://doi.org/10.3847/1538-4357/ab2c03

Morbidelli, A., Tsiganis, K., Crida, A., Levison, H.F., Gomes, R., 2007. Dynamics of the Giant Planets of the solar system in the Gaseous Protoplanetary Disk and Their Relationship to the Current Orbital Architecture. Astron. J. 134, 1790. https://doi.org/10.1086/521705

Morishima, R., Stadel, J., Moore, B., 2010. From planetesimals to terrestrial planets: N-body simulations including the effects of nebular gas and giant planets. Icarus 207, 517–535. https://doi.org/10.1016/j.icarus.2009.11.038

Mukhopadhyay, S., 2012. Early differentiation and volatile accretion recorded in deep-mantle neon and xenon. Nature 486, 101–104. https://doi.org/10.1038/nature11141

Nagasawa, M., Lin, D.N.C., Thommes, E., 2005. Dynamical Shake-up of Planetary Systems. I. Embryo Trapping and Induced Collisions by the Sweeping Secular Resonance and Embryo-Disk Tidal Interaction. Astrophys. J. 635, 578. https://doi.org/10.1086/497386

Nagasawa, M., Tanaka, H., Ida, S., 2000. Orbital Evolution of Asteroids during Depletion of the Solar Nebula. Astron. J. 119, 1480. https://doi.org/10.1086/301246

Nemchin, A., Timms, N., Pidgeon, R., Geisler, T., Reddy, S., Meyer, C., 2009. Timing of crystallization of the lunar magma ocean constrained by the oldest zircon. Nat. Geosci. 2, 133–136. https://doi.org/10.1038/ngeo417

Nesvorný, D., 2011. YOUNG SOLAR SYSTEM\textquotesingles FIFTH GIANT PLANET? Astrophys. J. 742, L22. https://doi.org/10.1088/2041-8205/742/2/L22

Nesvorný, D., Morbidelli, A., 2012. STATISTICAL STUDY OF THE EARLY SOLAR SYSTEM'S INSTABILITY WITH FOUR, FIVE, AND SIX GIANT PLANETS. Astron. J. 144, 117. https://doi.org/10.1088/0004-6256/144/4/117

Nesvorný, D., Roig, F.V., Deienno, R., 2021. The Role of Early Giant-planet Instability in Terrestrial Planet Formation. Astron. J. 161, 50. https://doi.org/10.3847/1538-3881/abc8ef

O'Brien, D.P., Morbidelli, A., Levison, H.F., 2006. Terrestrial planet formation with strong dynamical friction. Icarus 184, 39–58. https://doi.org/10.1016/j.icarus.2006.04.005

Ozima, M., Podosek, F.A., 1999. Formation age of Earth from 129I/127I and 244Pu/238U systematics and the missing Xe. J. Geophys. Res. Solid Earth 104, 25493–25499. https://doi.org/10.1029/1999JB900257

Pierens, A., Raymond, S.N., Nesvorny, D., Morbidelli, A., 2014. OUTWARD MIGRATION OF JUPITER AND SATURN IN 3:2 OR 2:1 RESONANCE IN RADIATIVE DISKS: IMPLICATIONS FOR THE GRAND TACK AND NICE MODELS. Astrophys. J. 795, L11. https://doi.org/10.1088/2041-8205/795/1/L11

Quintana, E.V., Barclay, T., Borucki, W.J., Rowe, J.F., Chambers, J.E., 2016. THE FREQUENCY OF GIANT IMPACTS ON EARTH-LIKE WORLDS. Astrophys. J. 821, 126. https://doi.org/10.3847/0004-637X/821/2/126

Raymond, S.N., Izidoro, A., 2017. The empty primordial asteroid belt. Sci. Adv. 3, e1701138. https://doi.org/10.1126/sciadv.1701138

Raymond, S.N., O'Brien, D.P., Morbidelli, A., Kaib, N.A., 2009. Building the terrestrial planets: Constrained accretion in the inner solar system. Icarus 203, 644–662. https://doi.org/10.1016/j.icarus.2009.05.016

Raymond, S.N., Quinn, T., Lunine, J.I., 2006. High-resolution simulations of the final assembly of Earth-like planets I. Terrestrial accretion and dynamics. Icarus 183, 265–282. https://doi.org/10.1016/j.icarus.2006.03.011

Raymond, S.N., Schlichting, H.E., Hersant, F., Selsis, F., 2013. Dynamical and collisional constraints on a stochastic late veneer on the terrestrial planets. Icarus 226, 671–681. https://doi.org/10.1016/j.icarus.2013.06.019

Ribeiro, R. de S., Morbidelli, A., Raymond, S.N., Izidoro, A., Gomes, R., Vieira Neto, E., 2020. Dynamical evidence for an early giant planet instability. Icarus 339, 113605. https://doi.org/10.1016/j.icarus.2019.113605

Rosenblatt, P., Charnoz, S., Dunseath, K.M., Terao-Dunseath, M., Trinh, A., Hyodo, R., Genda, H., Toupin, S., 2016. Accretion of Phobos and Deimos in an extended debris disc stirred by transient moons. Nat. Geosci. 9, 581–583. https://doi.org/10.1038/ngeo2742

Rubie, D.C., Frost, D.J., Mann, U., Asahara, Y., Nimmo, F., Tsuno, K., Kegler, P., Holzheid, A., Palme, H., 2011. Heterogeneous accretion, composition and core–mantle differentiation of the Earth.

Earth Planet. Sci. Lett. 301, 31–42. https://doi.org/10.1016/j.epsl.2010.11.030

Rubie, D.C., Jacobson, S.A., Morbidelli, A., O'Brien, D.P., Young, E.D., de Vries, J., Nimmo, F., Palme, H., Frost, D.J., 2015. Accretion and differentiation of the terrestrial planets with implications for the compositions of early-formed solar system bodies and accretion of water. Icarus 248, 89–108. https://doi.org/10.1016/j.icarus.2014.10.015

Rudge, J.F., Kleine, T., Bourdon, B., 2010. Broad bounds on Earth's accretion and core formation constrained by geochemical models. Nat. Geosci. 3, 439–443. https://doi.org/10.1038/ngeo872

Sanloup, C., Jambon, A., Gillet, P., 1999. A simple chondritic model of Mars. Phys. Earth Planet. Inter. 112, 43–54. https://doi.org/10.1016/S0031-9201(98)00175-7

Schiller, M., Bizzarro, M., Fernandes, V.A., 2018. Isotopic evolution of the protoplanetary disk and the building blocks of Earth and the Moon. Nature 555, 507–510. https://doi.org/10.1038/nature25990

Sleep, N.H., 2016. Asteroid bombardment and the core of Theia as possible sources for the Earth's late veneer component. Geochem. Geophys. Geosystems 17, 2623–2642. https://doi.org/10.1002/2016GC006305

Sleep, N.H., Zahnle, K.J., Kasting, J.F., Morowitz, H.J., 1989. Annihilation of ecosystems by large asteroid impacts on the early Earth. Nature 342, 139–142. https://doi.org/10.1038/342139a0

Staudacher, T., Allègre, C.J., 1982. Terrestrial xenology. Earth Planet. Sci. Lett. 60, 389–406. https://doi.org/10.1016/0012-821X(82)90075-9

Svetsov, V., 2011. Cratering erosion of planetary embryos. Icarus 214, 316–326. https://doi.org/10.1016/j.icarus.2011.04.026

Swindle, T.D., Caffee, M.W., Hohenberg, C.M., Taylor, S.R., 1986. I-Pu-Xe dating and the relative ages of the earth and moon. Proc. Conf. Kona HI Lunar Planet. Inst. Houst. TX Pubs 331–357.

Tanaka, H., Takeuchi, T., Ward, W.R., 2002. Three-Dimensional Interaction between a Planet and an Isothermal Gaseous Disk. I. Corotation and Lindblad Torques and Planet Migration. Astrophys. J. 565, 1257. https://doi.org/10.1086/324713

Tanaka, H., Ward, W.R., 2004. Three-dimensional Interaction between a Planet and an Isothermal Gaseous Disk. II. Eccentricity Waves and Bending Waves. Astrophys. J. 602, 388. https://doi.org/10.1086/380992

Tang, H., Dauphas, N., 2014. 60Fe–60Ni chronology of core formation in Mars. Earth Planet. Sci. Lett. 390, 264–274. https://doi.org/10.1016/j.epsl.2014.01.005

Taylor, D.J., McKeegan, K.D., Harrison, T.M., 2009. Lu–Hf zircon evidence for rapid lunar differentiation. Earth Planet. Sci. Lett. 279, 157–164. https://doi.org/10.1016/j.epsl.2008.12.030

Thiemens, M.M., Sprung, P., Fonseca, R.O.C., Leitzke, F.P., Münker, C., 2019. Early Moon formation inferred from hafnium–tungsten systematics. Nat. Geosci. 12, 696–700. https://doi.org/10.1038/s41561-019-0398-3

Thommes, E., Nagasawa, M., Lin, D.N.C., 2008. Dynamical Shake-up of Planetary Systems. II.N-Body Simulations of solar system Terrestrial Planet Formation Induced by Secular Resonance Sweeping. Astrophys. J. 676, 728–739. https://doi.org/10.1086/526408

Timpe, M.L., Han Veiga, M., Knabenhans, M., Stadel, J., Marelli, S., 2020. Machine learning applied to simulations of collisions between rotating, differentiated planets. Comput. Astrophys. Cosmol. 7, 2. https://doi.org/10.1186/s40668-020-00034-6

Trinquier, A., Birck, J.-L., Allègre, C.J., 2007. Widespread 54Cr Heterogeneity in the Inner solar system. Astrophys. J. 655, 1179. https://doi.org/10.1086/510360

Trinquier, A., Elliott, T., Ulfbeck, D., Coath, C., Krot, A.N., Bizzarro, M., 2009. Origin of Nucleosynthetic Isotope Heterogeneity in the Solar Protoplanetary Disk. Science 324, 374–376. https://doi.org/10.1126/science.1168221

Tsiganis, K., Gomes, R., Morbidelli, A., Levison, H.F., 2005. Origin of the orbital architecture of the giant planets of the solar system. Nature 435, 459–461. https://doi.org/10.1038/nature03539

Vorobyov, E.I., Elbakyan, V.G., 2018. Gravitational fragmentation and formation of giant protoplanets on orbits of tens of au. Astron. Astrophys. 618, A7. https://doi.org/10.1051/0004-6361/201833226

Walker, R.J., 2009. Highly siderophile elements in the Earth, Moon and Mars: Update and implications for planetary accretion and differentiation. Geochemistry 69, 101–125. https://doi.org/10.1016/j.chemer.2008.10.001

Walsh, K.J., Levison, H.F., 2019. Planetesimals to terrestrial planets: Collisional evolution amidst a dissipating gas disk. Icarus 329, 88–100. https://doi.org/10.1016/j.icarus.2019.03.031

Walsh, K.J., Morbidelli, A., Raymond, S.N., O'Brien, D.P., Mandell, A.M., 2011. A low mass for Mars from Jupiter's early gas-driven migration. Nature 475, 206–209. https://doi.org/10.1038/nature10201

Wetherill, G.W., Stewart, G.R., 1989. Accumulation of a swarm of small planetesimals. Icarus 77, 330–357. https://doi.org/10.1016/0019-1035(89)90093-6

Wisdom, J., Holman, M., 1991. Symplectic maps for the n-body problem. Astron. J. 102, 1528–1538. https://doi.org/10.1086/115978

Woo, J.M.Y., Brasser, R., Matsumura, S., Mojzsis, S.J., Ida, S., 2018. The curious case of Mars' formation. Astron. Astrophys. 617, A17. https://doi.org/10.1051/0004-6361/201833148

Woo, J.M.Y., Genda, H., Brasser, R., Mojzsis, S.J., 2019. Mars in the aftermath of a colossal impact. Icarus 333, 87–95. https://doi.org/10.1016/j.icarus.2019.05.015

Woo, J.M.Y., Grimm, S., Brasser, R., Stadel, J., 2021a. Growing Mars fast: High-resolution GPU simulations of embryo formation. Icarus 359, 114305. https://doi.org/10.1016/j.icarus.2021.114305

Woo, J.M.Y., Stadel, J., Grimm, S., Brasser, R., 2021b. Mars' Formation Can Constrain the Primordial Orbits of the Gas Giants. Astrophys. J. Lett. 910, L16. https://doi.org/10.3847/2041-8213/abed56

Yamakawa, A., Yamashita, K., Makishima, A., Nakamura, E., 2010. CHROMIUM ISOTOPE SYSTEMATICS OF ACHONDRITES: CHRONOLOGY AND ISOTOPIC HETEROGENEITY OF THE INNER SOLAR SYSTEM BODIES. Astrophys. J. 720, 150–154. https://doi.org/10.1088/0004-637X/720/1/150

Yin, Q., Jacobsen, S.B., Yamashita, K., Blichert-Toft, J., Télouk, P., Albarède, F., 2002. A short timescale for terrestrial planet formation from Hf–W chronometry of meteorites. Nature 418, 949–952. https://doi.org/10.1038/nature00995

Yu, G., Jacobsen, S.B., 2011. Fast accretion of the Earth with a late Moon-forming giant impact. Proc. Natl. Acad. Sci. 108, 17604–17609. https://doi.org/10.1073/pnas.1108544108